%% file: Scalar_resonance_in_a_top_partner_model.tex
\documentclass[12pt]{article}
\pdfoutput=1

\usepackage[DIV13]{typearea}
\usepackage{indentfirst}
\usepackage{amsmath}
\usepackage{amsfonts}
\usepackage{mathrsfs}
\usepackage{amssymb}
\usepackage{mathtools}
\usepackage{epsfig}
\usepackage{graphicx}
\usepackage{subfigure}
\usepackage{slashed}
\usepackage{multicol}
\usepackage[usenames,dvipsnames]{color}
\usepackage{cite}
\RequirePackage[colorlinks=true,urlcolor=blue,anchorcolor=blue,citecolor=blue,filecolor=blue,
               linkcolor=blue,menucolor=blue,linktocpage=true,pdfproducer=medialab]{hyperref}
\usepackage{cancel}

\usepackage{feynmp}
\makeatletter
\def\endfmffile{%
  \fmfcmd{\p@rcent\space the end.^^J%
          end.^^J%
          endinput;}%
  \if@fmfio
    \immediate\closeout\@outfmf
  \fi
  \ifnum\pdfshellescape=\@ne
    \immediate\write18{mpost \thefmffile}%
  \fi}
\makeatother

\usepackage{lscape}
\usepackage{multirow}
\usepackage{array}
\usepackage{pdfpages}
\usepackage{fancyhdr}
\usepackage{pifont}

\usepackage[left=.9in, right=.9in]{geometry}

\newcommand{\email}[1]{\href{mailto:#1}{\tt #1}}

\numberwithin{equation}{section}
\newcommand{\LL}{\mathscr{L}}
\newcommand{\cG}{\mathscr{G}}
\def\cB{{\cal B}}

\def\cG{{\cal G}}

\def\cJ{{\cal J}}

\def\cO{{\cal O}}

\def\cT{{\cal T}}

\def\Tr{{\rm Tr}}

\def\cw{c_{\textrm w}}
\def\sw{s_{\textrm w}}

\def\be{\begin{equation}}
\def\ee{\end{equation}}
\def\beq{\begin{equation}}
\def\eeq{\end{equation}}
\def\bc{\begin{center}}
\def\ec{\end{center}}
\def\bea{\begin{eqnarray}}
\def\eea{\end{eqnarray}}

\def\bry{\begin{array}}
\def\ery{\end{array}}

\def\nt{\noindent}

\newcommand{\hc}{\text{h.c.}}
\def\fA{\mbox{{\bf M$_{4+5}$}}}
\def\fB{\mbox{{\bf M$_{4+14}$}}}
\def\oA{\mbox{{\bf M$_{1+5}$}}}
\def\oB{\mbox{{\bf M$_{1+14}$}}}

\def\lra#1{\overset{\text{\scriptsize$\leftrightarrow$}}{#1}}

\def\fourpletL{\Psi_{\textbf{4}L}}
\def\fourpletR{\Psi_{\textbf{4}R}}
\def\fourplet{\Psi_{\bf{4}}}

\def\Barfourplet{\overline{\Psi}_{\bf{4}}}

\def\singletL{\Psi_{\textbf{1}L}}
\def\singletR{\Psi_{\textbf{1}R}}
\def\singlet{\Psi_{\bf{1}}}

\def\Barsinglet{\overline{\Psi}_{\bf{1}}}

\def\5qplet{q_L^{\mathbf{5}}}
\def\qBar5plet{\overline{q}^{\bf{5}}_L}
\def\14qplet{q_L^{\bf{14}}}
\def\q14Barplet{\overline{q}^{\bf{14}}_L}

\def\u5plet{u_R^{\bf 5}}
\def\uBar5plet{\overline{u}^{\bf{5}}_R}
\def\usinglet{u_R^{\bf 1}}

\def\d5plet{d_R^{\bf 5}}
\def\dBar5plet{\overline{d}^{\bf{5}}_R}

\def\Xtt{{X_{\hspace{-0.09em}\mbox{\scriptsize2}\hspace{-0.06em}{\raisebox{0.1em}{\tiny\slash}}\hspace{-0.06em}\mbox{\scriptsize3}}}}
\def\Xft{{X_{\hspace{-0.09em}\mbox{\scriptsize5}\hspace{-0.06em}{\raisebox{0.1em}{\tiny\slash}}\hspace{-0.06em}\mbox{\scriptsize3}}}}
\def\Tt{\widetilde{T}}

\newcommand{\cJmuuq}{\cJ^\mu_q}
\newcommand{\cJmuuu}{\cJ^\mu_u}

\newcommand{\cJmuupsi}{\cJ^\mu_\psi}
\newcommand{\cJmuuqpsi}{\cJ^\mu_{q\psi}}
\newcommand{\cJmuuupsi}{\cJ^\mu_{u\psi}}

\newcommand{\aichi}{\alpha^\chi_i}

\newcommand{\eg}{e.g.\,\,}
\newcommand{\ie}{i.e.\,\,}

\newcommand{\hhref}[1]{\href{http://arxiv.org/abs/#1}{arXiv:#1}}

\begin{document}
\begin{titlepage}
\vspace*{-1cm}
\phantom{hep-ph/***} 
\vskip 1cm
\begin{center}
\mathversion{bold}
{\LARGE\bf Scalar resonance in a top partner model}\\
\mathversion{normal}
\vskip .3cm
\end{center}
\vskip 0.5  cm
\begin{center}
{\large Sebasti\'an Norero}~$^{a)}$,
{\large Juan Yepes}~$^{a)}$ and
{\large Alfonso Zerwekh}~$^{a)}$
\\
\vskip .7cm
{\footnotesize
$^{a)}$~
\emph{Department of Physics and Centro Cient\'{i}fico-Tecnol\'{o}gico de Valpara\'{i}so\\
Universidad T\'{e}cnica Federico Santa Mar\'{i}a, Valpara\'{i}so, Chile}\\
\vskip .1cm
\vskip .3cm
\begin{minipage}[l]{.9\textwidth}
\begin{center} 
\textit{E-mail:} 
\email{sebastian.norero@alumnos.usm.cl},
\email{juan.yepes@usm.cl},
\email{alfonso.zerwekh@usm.cl}
\end{center}
\end{minipage}
}
\end{center}
\vskip 0.5cm
\begin{abstract}
\nt The phenomenology entailed by a scalar resonance in a top partner model is analysed here in a $SO(5)$ Composite Higgs formalism. Heavy scalar resonances production and their decays modes are explored along a benchmark resonance mass range. The production of single-double partner final states has been scanned along the partner mass scale. QCD drives such production, as well as the SM gauge, Higgs, plus the intermediation of the scalar resonance. Non-zero contributions are induced as long as extra fermion-resonance effects are included. Finally, we have excluded regions of the parameter spaces underlying our framework by imposing the recent LHC searches for vector-like quarks production in $pp$-collisions at 13 TeV. Substantial reduction of the allowed regions occurs if extra fermion-resonance effects are accounted for, leading us to test the involved parametric dependence in the shed light of new matter interactions.
\end{abstract}
\end{titlepage}
\setcounter{footnote}{0}

\tableofcontents

%
%

\newpage

\section{Introduction}

\nt The UV insensitivity of the Higgs boson mass is assured if new particles canceling the contributions from the the Standard Model (SM) top quarks play a role at higher energies. The Higgs discovery at the LHC~\cite{Aad:2012tfa,Chatrchyan:2012xdj} has prompt us to explore new feasible beyond SM (BSM) frameworks, aimed at healing the UV sensitivity of the Higgs mass. Some BSM frameworks propose new states, currently called \emph{top partners}, as in the case of the scalar quarks for the well known supersymmetry, or vector-like fermions~\cite{Aguilar-Saavedra:2013qpa,Panizzi:2014dwa} as in composite Higgs models (CHMs)~\cite{Kaplan:1983fs,Kaplan:1983sm,Georgi:1984ef,Banks:1984gj,Georgi:1984af,Dugan:1984hq,Contino:2003ve,Agashe:2004rs,Contino:2010rs}. Vector-like quarks are hypothetical spin-1/2 particles with left- and right-handed components behaving in the same way under the SM symmetries, and extensively analysed in the CHM literature~\cite{Contino:2006qr,Matsedonskyi:2012ym,Marzocca:2012zn,Pomarol:2012qf,Redi:2012ha,Panico:2012uw}. Suplementarily, exotic spin-0 and spin-1 resonances at the TeV scale have been considered in such models, impacting on the pseudo NG bosons (PNGBs) scattering, and then on the high-energy vector boson scattering~\cite{Contino:2011np}.

Guided us by the challenging scenario of weak scale naturalness, we explore the low energy effects from the interplay among: elementary, composite partners and spin-0 resonances in a $SO(5)$ CHM. We encode such interactions via derivative couplings of the scalar resonance $\eta$, here assumed to be a singlet of $SO(4)$, with a complete set of $SO(5)$-invariant fermionic currents presented in this analysis. Such invariants cover all the structures built upon the SM elementary sector together with the top partners embedded in the unbroken $SO(4)$, concretely, a fourplet $\fourplet$ and a singlet $\singlet$, naturally sourced by the decomposition rule ${\bf 5}={\bf 4}+{\bf 1}$ under $SO(4)$,  encoded through 
\be
\fourplet={1\over \sqrt{2}}\left(\begin{matrix}
i\cB-i\Xft\\
\cB+\Xft\\
i\cT+i\Xtt\\
-\cT+\Xtt
\end{matrix}\right),\qquad\qquad \singlet=\widetilde{\cT}\,.
\label{fourplet-singlet}
\ee

\nt Two doublets $(\cT,\cB)$ and $(\Xft,\Xtt)$ compose the fourplet $\fourplet$. The former has hypercharge $1/6$ and the same quantum numbers as the SM quark doublet, whilst the latter has hypercharge $7/6$, containing a state of exotic charge $5/3$ plus another top-like quark $\Xtt$. The singlet representation $\singlet$ entails only one exotic top-like state, denoted in here as $\widetilde{\cT}$. On the other hand, the elementary sector will be shaped according to the partial compositeness mechanism via the Goldstone symmetry breaking Lagrangian
\be
{\cal L}_{\text{mix}}= \sum_q y\,\bar q\,\cO_q .
\label{UV-mix}
\ee

\nt Two choices for the elementary sector embeddings are determined by the $SO(5)$  strong sector operators $\cO_q$: either as a fundamental ${\bf 5}$ or ${\bf 14}$ representation. In the former scenario, both fermion chiralities have elementary representatives coupled to the strong sector through ${\bf 5}$-plets
\be
\hspace*{0.5cm}
\5qplet={1\over \sqrt{2}}\left(
i d_L,\,\,
 d_L,\,\,
i u_L,\,\,
- u_L,\,\,
0
\right)^T,
\qquad\quad
\u5plet = \left(
0,\,\,
0,\,\,
0,\,\,
0,\,\,
u_R
\right)^T,
\label{emb}
\ee

\nt whereas in the latter the right-handed $q$ quark enters as a totally composite state arising itself from the operator ${\cal Q}_q$ at low energies with the fields
\be
\hspace*{0.5cm}
\14qplet={1\over \sqrt{2}}\left(\begin{matrix}
0 & 0 & 0 & 0 & i d_L\\
0 & 0 & 0 & 0 & d_L\\
0 & 0 & 0 & 0 & i u_L\\
0 & 0 & 0 & 0 & -u_L\\
i d_L & d_L & i u_L & -u_L &0\\
\end{matrix}\right),
\quad
\usinglet\,.
\label{emb}
\ee

\nt  In both cases the representations $q_L$ and $u_R$ have the same $X$-charge $2/3$, allowing to reproduce the correct electric charge of the top. The doublet $q^T_L=(u_L,d_L)$ has an isospin $T_R^3=-1/2$, providing thus a protection from large deformations of the $b_L$-couplings~\cite{Agashe:2006at,Mrazek:2011iu}. Four models will be framed following the previous matter content, each of them generically described at the Lagrangian level through
\be
\LL = \LL_{\text{elem}}\,\,+\,\,\,\LL_{\text{comp}}\,\,+\,\,\,\LL_{\text{mix}}.
\label{Lagrangian}
\ee

\nt Scenario that will be coupled later on to the scalar resonance $\eta$, described here as a singlet $\eta=(\mathbf{1},\mathbf{1})$ of $SU(2)_L\times SU(2)_R$ that have been considered in CHMs~\cite{Contino:2011np}. In the next sections all the Lagrangians will be provided.

Top quark physics at CHMs have been extensively studied ~\cite{Marzocca:2012zn,DeSimone:2012fs,ewpt}, with general flavour physics analyses~\cite{Barbieri:2012tu,Redi:2012uj,KerenZur:2012fr}, considered in the context of top partner sectors~\cite{Matsedonskyi:2014iha}, whilst spin-0 and spin-1 resonances have been considered in CHMs~\cite{Contino:2011np} with updated analysis~\cite{Hernandez:2015xka}. Our discussion will be based on the previous studies~\cite{Matsedonskyi:2014iha,DeSimone:2012fs}, recently armed with heavy triplet vector resonances in~\cite{Yepes:2017pjr}, and whose phenomenology signals have been thoroughly explored in~\cite{Yepes:2018dlw}, but extended in this work up to a simple approach for effective top partners-scalar resonances interplay instead. The scalar resonance production and their decays modes are exhaustively explored along a viable range for the resonance mass $M_\eta$ and for a given setting of the parameters in our model. Likewise, the production of single-double partner final states has been scanned along the partner mass scale $M_\Psi$, and they turn out to be controlled by a set of model-dependent couplings here provided. Double production is QCD-driven, as well as SM gauge, Higgs, and $\eta$-mediated. Non-zero parametric-dependent modifications are induced as long as extra fermion-resonance effects are considered.

Finally, we have been imposed the recent LHC searches for vector-like quarks production in $pp$-collisions at 13 TeV~\cite{Sirunyan:2017pks} in order to exclude regions of the parameter spaces underlying our model. Specifically, we explore the allowed regions by bounding the decays $\cT\to Wb$ and $\widetilde{\cT}\to Wb$ according with the latest experimental limits. Substantial reduction of the permitted regions occurs when the extra fermion-resonance couplings treated here are accounted for. The sensitivity of the parametric dependence is thus tested in the presence of new exotic matter interactions.

This manuscript is divided in: introduction of the PNGB's at the assumed CHM, scalar resonance sector and its generic interplay with the elementary-composite sector in Section-\ref{Interplay}. Scalar resonances production and their decays in Section~\ref{Resonance-production-decay}. Top partners production mechanism are introduced in Section~\ref{Top-partner-production} and detailed discussed in~\ref{Double-Partner-production}-\ref{Single-Partner-production}. The latest LHC searches on vector-like quark production are translated into parameter spaces associated to our models in Section~\ref{Some-parameter-spaces}. The impact of the additional fermion-resonance interactions is extensively studied along the text. A summary is presented in Section~\ref{Summary}.

\section{Set-up formalism}
\label{Interplay}

\nt The \emph{composite sector} entails a composite Higgs boson and other composite resonances all described by the CCWZ formalism~\cite{ccwz}. It postulates the Higgs as a PNGB of the minimal global symmetry $\cG=SO(5)$~\cite{Agashe:2004rs} and spontaneously broken to $SO(4)$ by the strong sector at the scale $f$. Such breaking generates four massless PNGBs, forming an $SU(2)_L$ Higgs doublet a posteriori. At this level the Higgs is exactly massless unless the strong sector is coupled to some source of an explicit $\cal G$-breaking. An additional $U(1)_X$ factor is introduced in order to restore the proper SM hypercharge $Y=T_R^3+X$, then $\cG=SO(5)\times U(1)_X$. The PNGBs enter through the  $5\times 5$ Goldstone matrix
\be
U=\exp\left[i \frac{\sqrt{2}}{f}\,\Pi^i\,T^i\right]=\,
\left(\begin{array}{ccccc}
& & \vspace{-3mm}& & \\
 & \mathbb{I}_{3} & & &  \\
  & & \vspace{-3mm}& &  \\ 
   & & & \cos \frac{h + \langle h \rangle}{f} & \sin \frac{h + \langle h \rangle}{f} \\
    & & & -\sin \frac{h + \langle h \rangle}{f} & \cos \frac{h + \langle h \rangle}{f} 
\end{array}\right)\,,
\label{GB-matrix}
\ee

\nt where $T^i$ are the coset $SO(5)/SO(4)$-generators, whilst $\Pi^i$ and $f$ are the PNGB fields and the decay constant respectively. Henceforth $T^i$ will stand for the coset generators, while $T^a$  for the unbroken ones, all them defined in~Appendix~\ref{CCWZ}.  

Additionally, the \emph{elementary sector}, containing copies of all the SM field sector except for the Higgs transforming under the SM gauge symmetry group ${\cal G}_{\text{SM}} \subset {\cal G}$. This sector is not $\cal G$ invariant, therefore the one-loop effective potential triggered by the elementary-composite interactions allows the Higgs to pick a mass, fixing thus its vacuum expectation value (VEV) in a ${\cal G}_{\text{SM}}$-breaking direction. The unbroken $SO(4)\times U(1)_X$ contains the SM  symmetry ${\cal G}_{\text{SM}}=SU(2)_L \times U(1)_Y$ whose breaking will be triggered via a non-zero Higgs VEV $\langle h \rangle \simeq v=246$~GeV, measuring together with the $SO(5)$ breaking scale $f$ the degree of tuning of the scalar potential through the ratio~\cite{Agashe:2004rs}
\be
\xi=\frac{v^2}{f^2}.
\label{Xi}
\ee

\nt $\xi$ controls the low energies SM departures and it cannot be too large. Generically, the value of $f$ must be large to suppress NP effects, but not too far from $v$ to maintain a tolerable tuning. Stringent constraints on $\xi$ have been reported following the current 95\% combined limit from direct production of either vectorial charged $\rho^\pm$, or the neutral $\rho^0$ at the LHC~\cite{Contino:2013gna} (see~\cite{Yepes:2017pjr,Yepes:2018dlw} for a detailed discussion). Those limits allow $\xi\sim 0.02$, or even smaller, for a vector resonance mass $M_\rho\sim 2$ TeV. Such small values might be directly tested through single Higgs production at the LHC, reaching larger precision via double Higgs processes at CLIC, and should be compared with indirect bounds from EW precision data. For the present work we will test $\xi=$0.1-0.2, as they are compatible with the latter EWPT bounds, and with the vector resonance direct production bounds at LHC, as well as the expected single Higgs production at the LHC, and the double Higgs production at CLIC. In addtion, those values are inside the domain of validity of the scenario, $g_\rho<4\pi$ and they will be assumed henceforth. 

We  will cover all the possible couplings arising out the interplay among the top partners sector and the composite operators sourced from the strong regime.  The $SO(5)$-invariance will prescribe the generic Lagrangian
\be
\LL_{\text{int}}=\LL_{\bf M}\,\,+\,\,\LL_\eta\,\,+\,\,\LL_{\bf M\,+\,\eta}.
\label{Interplay} 
\ee

\nt $\bf M$ labels each one of the models emerging from the assumed fermionic matter content
\be
\bf M=\bf M_{\Psi +q}=\{\fA,\,\fB,\,\oA,\,\oB\}.
\label{Models}
\ee

\nt $\LL_{\bf M}$ is generically encoded by~\eqref{Lagrangian}, whilst $\LL_\eta$ describes the scalar sector and its interaction with the gauge fields. We are concerned here only to the case of resonances transforming under $SO(4)$. According to the rule $\mathbf{4}\times \mathbf{4} = \mathbf{1}+ \mathbf{6} +\mathbf{9}$ the resonance can therefore be encoded by one of the $SU(2)_L\times SU(2)_R$-representations $(\mathbf{1},\mathbf{1}) + (\mathbf{3},\mathbf{1}) + (\mathbf{1},\mathbf{3}) + (\mathbf{3},\mathbf{3})$. For the work undertaken in here, only the spin-0 resonances $\eta=(\mathbf{1},\mathbf{1})$ will be analysed. The spin-1 heavy triplet case has been recently considered in~\cite{Yepes:2017pjr,Yepes:2018dlw}. $\LL_\eta$ is given then as\footnote{Analogous interactions for a heavy scalar state have been treated in the context of heavy ``radial” excitation, parameterized by the NGBs of the \emph{global Higgs} models~\cite{Fichet:2016xvs,Fichet:2016xpw}.} 
\begin{equation}
{\cal L}_\eta = \frac{1}{2} \left( \partial_\mu \eta \right)^2 - \frac{1}{2} M_\eta^2 \eta^2 
+\frac{f^2}{4} \left( 2 a_\eta \frac{\eta}{f} + b_\eta \frac{\eta^2}{f^2} \right)  \Tr\left[ d_\mu d^\mu\right] \, ,
\label{eta-Lagrangian}
\end{equation}

\nt where non-zero parameters $a_\eta$ and $ b_\eta$ allow for $\eta\pi^2$ and $\eta^2\pi^2$ interactions relevant for $W$ and $Z$-pair scattering. Indeed, for $a_\eta = b_\eta =1$ the Lagrangian~\eqref{eta-Lagrangian} describes a linear sigma model, with $\eta$ and the $SO(5)/SO(4)$ NG bosons fitting together in a fundamental (linearly-transforming) representation of $SO(5)$. For that particular choice all the scattering amplitudes are perturbatively unitarized if $\eta$ is lighter than the cutoff (see~\cite{Contino:2011np} and references therein).

The third Lagrangian in~\eqref{Interplay} encodes fermion currents coupled to the scalar resonance completley provided by the first time in here, and generically defined as
\be
\LL_{\bf M\,+\,\eta}=\frac{\alpha_i}{\sqrt{2}\,f}\,\mathcal{J}^\mu_i\partial_\mu\eta\,\,+\,\,\hc
\,,
\label{Derivative-couplings}
\ee

\nt with $i=\{q,\,\psi,\,q\psi,\,u\psi\}$, denoting thus all the possible currents constructable upon the elementary $q$, top partner $\psi$ and elementary-top partner sector $q\psi$ and $u\psi$. Generic coefficients $\aichi$ have been introduced and are correspondingly weighting each one of the fermion currents defined later on.

\subsection{$\fA$ and $\oA$ coupled to $\eta$}

\nt The leading order Lagrangian for the $\bf{5}$-elementary fermions and the one describing both of the top partners $\fourplet$ and $\singlet$, introduced in $\LL_{\text{comp}}$~\eqref{Lagrangian},  are given by
\be
\LL_{\text{elem}}= i\,\overline{q}_L \slashed{D}\,q_L\,\,+\,\,i\,\overline{u}_R\slashed{D}\,u_R,
\label{fA-oA-elem}
\ee
\be
\begin{aligned}
\LL_{\text{comp}}= i\,\Barfourplet\slashed{\nabla}\fourplet - M_{\bf{4}}\,\Barfourplet \fourplet\,+\,\left(\fourplet \leftrightarrow\singlet\right)\,+\,\frac{f^2}{4}d^2\,+\,\left(i\,c_{41}\, (\Barfourplet)^i \gamma^\mu d_\mu^i \singlet + {\rm h.c.}\right)
\label{fA-oA-comp}
\end{aligned}
\ee

\nt with $\nabla$ standing for $\nabla=\slashed{D}+i\slashed{e}$. Goldstone bosons kinetic terms are contained at the $d^2$-term, while the coefficient $c_{41}$ controls the strength of the interplaying fourplet-singlet partner term, and it is is expected to be order one by power counting~\cite{Giudice:2007fh}. The covariant derivatives through~\eqref{fA-oA-elem}-\eqref{fA-oA-comp}, together with the $d$ and $e$-symbols are defined in~\ref{CCWZ}. Finally, the mass terms mixing the elementary and top partners are described via 
\be
\begin{aligned}
\LL_{\text{mix}} =& y_L f \left(\qBar5plet\,U\right)_i \,\left(\fourpletR\right)^{i}\,\,+\,\,y_R f \left(\uBar5plet\,U\right)_i \,\left(\fourpletL\right)^{i}\,\,+\,\,\hc\,\,+\,\,,\\[5mm]
& + \tilde{y}_L f \left(\qBar5plet\,U\right)_5\singletR\,\,+\,\,\tilde{y}_R f \left(\uBar5plet\,U\right)_5\singletL\,\,+\,\,\hc
\label{fA-oA-mix}
\end{aligned}
\ee

\nt whereas the trilinear couplings fermion-fermion-scalar are encoded through

\be
\begin{aligned}
\LL_{\text{mix}-\eta} =& \Bigl[y_{q\psi}\left(\qBar5plet\,U\right)_i \,\left(\fourpletR\right)^{i}\,\,+\,\,y_{u\psi}\left(\uBar5plet\,U\right)_i \,\left(\fourpletL\right)^{i}\,\,+\,\,\hc\,\,+\,\,\\[5mm]
& + \tilde{y}_{q\psi} \left(\qBar5plet\,U\right)_5\singletR\,\,+\,\,\tilde{y}_{u\psi}\left(\uBar5plet\,U\right)_5\singletL\,\,+\,\,\hc\Bigr]\eta\,.
\label{fA-oA-mix-eta}
\end{aligned}
\ee

\nt Suitable $U$ insertions have been done in order to guarantee the non-linear $SO(5)$ invariance. The small mixings $y_{L(R)}$ and $\tilde{y}_{L(R)}$ trigger the Goldstone symmetry breaking,  providing thus a proper low Higgs mass. The latter Lagrangian entails partially composite $u^5_R$ and it gives rise to quark mass terms as well as trilinear couplings contributing to the single production of top partners. 

The set of fermion currents constructable for both of the models $\fA$ and $\oA$ are listed in Table~\ref{Fermion-currents-set} (left column). Altogether, the leading order composite and mixing Lagrangians contain eleven parameters $\{M_{\bf{4}},\,M_{\bf{1}},\,c_{41},\, y_{L(R)},\, \tilde{y}_{L(R)},\, y_{q\psi (u\psi)},\,\tilde{y}_{q\psi (u\psi)}\}$, aside from the Goldstone decay constant $f$. Six of them are arranged to reproduce the correct top mass plus the extra partner masses $\{m_{\Xft},\,m_{\Xtt},\,m_{\cT},\,m_{\cB},\,m_{\widetilde{\cT}}\}$. Their expressions are reported in Appendix~\ref{Physical-fermion-masses}.

\subsection{$\fB$ and $\oB$ coupled to $\eta$}

\nt The elementary kinetic Lagrangian corresponding to this model and the composite counterpart are straightforwardly written as
\be
\LL_{\text{elem}}= i\,\overline{q}_L \slashed{D}\,q_L\,,
\label{fB-oB-mass}
\ee
\be
\begin{aligned}
\LL_{\text{comp}}\quad\rightarrow\quad\LL_{\text{comp}}+i\,\overline{u}_R \slashed{D}\,u_R\,+\,\left(i\,c_{41}\, (\Barfourplet)^i \gamma^\mu d_\mu^i \singlet\,+\,i\,c_{4u}\, (\Barfourplet)^i \gamma^\mu d_\mu^i u_R + \hc\right),
\label{fB-oB-comp}
\end{aligned}
\ee

\nt where the $\LL_{\text{comp}}$ of~\eqref{fA-oA-comp} has been reshuffled in order to account for mixing terms $\fourplet$-$\singlet$ and the totally composite $u_R$ through the coefficients $c_{41}$ and $c_{4u}$ respectively. The elementary and top partners sector are mixed via
\be
\begin{aligned}
\LL_{\text{mix}}&= y_L\,f\left(U^t\,\q14Barplet\,U\right)_{i\,5} \,\left(\fourpletR\right)^{i}\,\,+\,\,\tilde{y}_L\,f\left(U^t\,\q14Barplet\,U\right)_{5\,5} \,\singletR\,\,+\,\,y_R\,f\left(U^t\,\q14Barplet\,U\right)_{5\,5} \,\usinglet\,\,+\,\,\hc
\label{fB-oB-mix}
\end{aligned}
\ee

\nt whereas the trilinear couplings fermion-fermion-scalar are encoded through
\be
\begin{aligned}
\LL_{\text{mix}-\eta}&= \Bigl[y_{q\psi}\,\left(U^t\,\q14Barplet\,U\right)_{i\,5} \,\left(\fourpletR\right)^{i}\,\,+\,\,\tilde{y}_{q\psi}\,\left(U^t\,\q14Barplet\,U\right)_{5\,5} \,\singletR\,\,+\,\,y_{qu}\,\left(U^t\,\q14Barplet\,U\right)_{5\,5} \,\usinglet\,\,+\,\,\hc\Bigr]\eta
\label{fB-oB-mix-eta}
\end{aligned}
\ee

\nt This case nvolves ten parameters $\{M_{\bf{4}},\,M_{\bf{1}},\,c_{41},\, c_{4u},\,y_{L(R)},\,\tilde{y}_L,\,y_{q\psi (qu)},\,\tilde{y}_{q\psi}\}$, five of them are arranged  to reproduce the correct top mass, plus extra four partner masses as the degeneracy $m_{\Xft}=m_{\Xtt}$ is implied and also manifested at the previous two models. Notice that a direct mixing coupling $u_R$ and $\singlet$ has been removed by a field redefinition. Table~\ref{Fermion-currents-set} lists the associated fermion currents (right column). For these models, the parametric dependence is shortened by one unity, as the number of implied currents is less than in the $\bf{5}$-models.

\begin{table}
\centering
\small{
\hspace*{-3mm}
\renewcommand{\arraystretch}{1.0}
\begin{tabular}{c||c}
\hline\hline
\\[-3mm]
$\fA$ & $\fB$
\\[0.5mm]
\hline\hline
\\[-2mm]
$\begin{array}{l}  
\cJmuuq=\,\qBar5plet\,\,\gamma^\mu\,\,\5qplet\\
\\[-1mm]
\cJmuuu=\,\uBar5plet\,\,\gamma^\mu\,\,\u5plet\\
\\[-1mm]
\cJmuupsi =\,\Barfourplet\,\gamma^\mu\,\,\fourplet
\\[3mm]
\cJmuuqpsi =\,\left(\qBar5plet\,U\right)_j\gamma^\mu\left(\fourpletL\right)^j\\[3mm]
\cJmuuupsi =\,\left(\uBar5plet\,U\right)_j\gamma^\mu\left(\fourpletR\right)^j
\\[3mm]
\end{array}$  &  
$\begin{array}{l}  
\cJmuuq =\,\left(U^T\,\q14Barplet\,U\right)_{5\,j}\,\gamma^\mu\,\left(U^T\,\14qplet\,U\right)_{j\,5}\\
\\[-1mm]
\cJmuupsi =\,\Barfourplet\,\gamma^\mu\,\,\fourplet\\[4mm]\cJmuuqpsi =\,\left(U^T\,\q14Barplet\,U\,\right)_{5\,j}\gamma^\mu\,\left(\fourpletL\right)^j\\[4mm]
\cJmuuu =\,\bar{u}_R\,\gamma^\mu\,\,u_R
\end{array}$\\
\hline\hline
\\[-4mm]
$\oA$ & $\oB$
\\[0.5mm]
\hline\hline
\\   
$\begin{array}{l}  
\cJmuuq =\,\qBar5plet\,\,\gamma^\mu\,\,\5qplet
\\[5mm]
\cJmuuu=\,\uBar5plet\,\,\gamma^\mu\,\,\u5plet
\\[5mm]
\cJmuupsi=\,\Barsinglet\,\,\gamma^\mu\,\,\singlet
\\[5mm]
\mathcal{J}^{\mu}_{q\psi}=\,\left(\qBar5plet\,U\right)^5\,\gamma^\mu\,\singletL\\[5mm]
\mathcal{J}^{\mu}_{u\psi} =\,\left(\uBar5plet\,U\right)^5\,\gamma^\mu\,\singletL\\[4mm]
\end{array}$  &  
$\begin{array}{l}  
\\[-6mm]
\cJmuuq =\,\left(U^T\,\q14Barplet\,U\right)_{5\,j}\,\gamma^\mu\,\left(U^T\,\14qplet\,U\right)_{j\,5}\\[4mm]
\cJmuupsi=\,\Barsinglet\,\,\gamma^\mu\,\,\singlet
\\[5mm]
\mathcal{J}^{\mu}_{q\psi} =\,\left(U^T\,\q14Barplet\,U\right)_{5\,5}\gamma^\mu\,\singletL\\[4mm]
\cJmuuupsi =\,\bar{u}_R\,\gamma^\mu\,\singletR
\end{array}$\\[2mm]  
\hline \hline
\end{tabular}
\caption{\sf Currents for all the models. }  
\label{Fermion-currents-set}
}
\end{table}


\section{Spin-0 production and decays}
\label{Resonance-production-decay}

\nt Concerning the resonance production, the role of spin-0 and spin-1 resonances on the PNGBs scattering were studied in~\cite{Contino:2011np}. Their experimental searches~\cite{ATLAS:2013jma} were explored for $\xi=0.1$ in~\cite{Contino:2013gna,Pappadopulo:2014qza,Greco:2014aza}, while the impact of heavy triplet resonances at the LHC in the final states $l^+l^-$ and $l\nu_l$ ($l=e,\mu$), $\tau^+\tau^-$, $jj$, $t\bar{t}$ as well as on the gauge and gauge-Higgs channels $WZ$, $WW$, $WH$ and $ZH$, has been analysed (see~\cite{Shu:2016exh,Shu:2015cxm,Zerwekh:2005wh} and references therein), constraining the vector resonance mass in the range 2.1-3 TeV. On the other hand, searches for Higgs-like bosons in the range 80-140 TeV decaying into long-lived exotic particles, have obtained no excess above the background expectation~\cite{Aaij:2016isa}. Likewise, searches for massive long-lived particles decaying semileptonically in the LHCb detector found no experimental evidences at the EW scale~\cite{Aaij:2016xmb}. Meanwhile, the searches performed by the ATLAS Collaboration for heavy scalar resonances decaying into $W W$ in the $e\nu \mu \nu$ final state, via $pp$ collisions at $\sqrt{s} = 13$ TeV with an integrated luminosity of $36.1\,\text{fb}^{-1}$, have revealed no significant excess of events beyond the SM background prediction in the mass range 200-5000 TeV~\cite{Aaboud:2017gsl}. In order to explore the feasibility and potentiality of our scenarios, a broader mass range will be explored in here. At the Lagrangian level, the scalar resonance production is induced by the effective charged-neutral interactions
\be
\hspace*{-1mm}
\LL_{ff\eta}= \sum_{f=u,d}\left[g_{ff \eta}\,{\bar f}\,f\,\eta\,\,+\,\,{\bar f}\,\slashed{\partial}\eta\left(g_{f_L f_L \eta}\,P_L+ g_{f_R f_R \eta}\,P_R\right)f\right].
\label{Scalar-Yukawa-derivative-couplings}
\ee

\begin{figure}
\begin{center}
\includegraphics[scale=0.8]{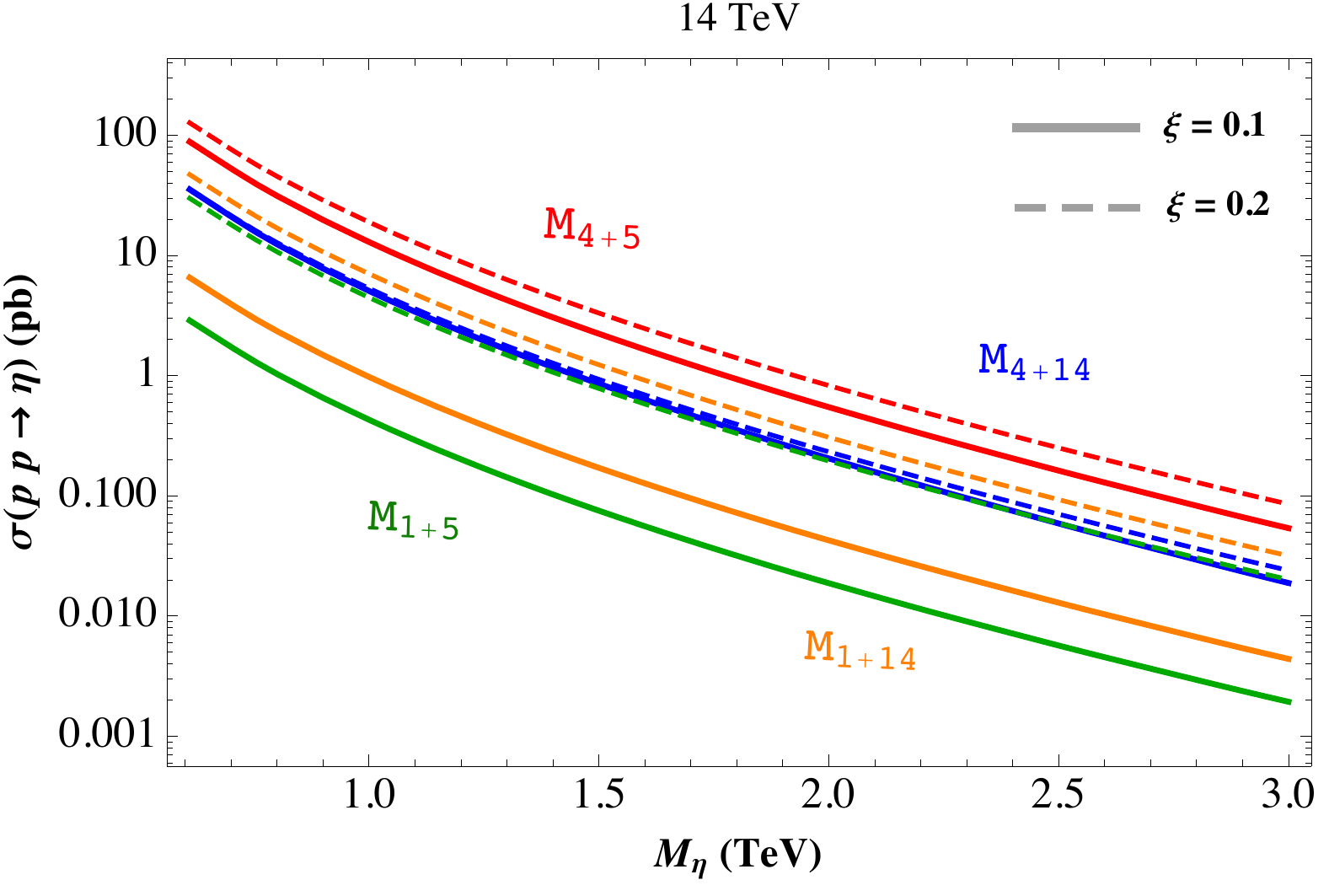}
\caption{\sf Production cross section for $\eta$ in all models at 14 TeV, for $\xi=0.1,\,0.2$ (thick and thin curves), and by setting $\alpha_i=1$.}
\label{eta-Production-cross-sections}
\end{center}
\end{figure}

\nt The couplings $g_{ff \eta}$ are depending on the coefficients in the Yukawa-scalar interactions, either of~\eqref{fA-oA-mix-eta} or~\eqref{fB-oB-mix-eta}, while $g_{f_L f_L \eta}$ and $g_{f_R f_R \eta}$ directly depends on the weighting coefficients $\alpha$ of~\eqref{Derivative-couplings}, as well as on fermion diagonalization effects (see Appendix~\ref{Effective-couplings}). The latter couplings are $f$-suppressed and they can be traded for slight modifications for the former $g_{ff \eta}$ via equations of motion. Therefore, the final Yukawa interactions are slightly affected by departures $\sim \alpha\,m_f/f$, with $m_f$ a given SM fermion mass. Consequently, such interactions turn out to be insensitive to the presence of the derivative couplings in~\eqref{Derivative-couplings}, although the setting $\alpha=1$ will be assumed for the computation of the scalar production. Associated production cross sections through the process $p\,p \to \eta$ are computed from the latter Lagrangians by using MadGraph 5. Fig.~\ref{eta-Production-cross-sections} displays all the spin-0 production cross sections as a functions of the parameter $M_\eta$ in the benchmark mass range $M_\eta \in [0.6,\,3]$\,TeV, for all the models at $\sqrt{s}=14$ TeV, and setting $\alpha=1$ for $\xi=0.1,\,0.2$. The resonance production is slightly altered when the fermion-resonance current interactions of~\eqref{Derivative-couplings} are included, therefore the situation $\alpha=0$ coincides with the one in Fig.~\ref{eta-Production-cross-sections}. Keeping the SM top quark mass at its experimental observed value requires the Yukawa couplings $y_L,\, y_R,\, \tilde{y}_L,\, \tilde{y}_R$ in~\eqref{fA-oA-mix} and \eqref{fB-oB-mix} to be properly set, either through its predicted value in~\eqref{Masses-expanded-5} or~\eqref{Masses-expanded-14} and by implementing relations in~\eqref{eta-parameters}. The scalar heavy resonances is predominantly yielded at the model $\fA$ as it can be seen from Fig.~\ref{eta-Production-cross-sections}. In addition, a higher $\xi$-value enhances all the productions, although at $\fB$ the production is slightly increased. Notice that whether the elementary fermions are $\bf{5}$ or $\bf{14}$-embeddings, the fourplet scenario favours higher production values rather than the singlet one. The scalar resonance is mainly yielded at $\fA$, reaching rough cross section values of $\sim 150$ pb (0.1 pb) at $M_\eta \sim 0.6$ TeV (3 TeV) for $\xi=0.2$. 

\begin{figure}
\begin{center}
\includegraphics[scale=0.6]{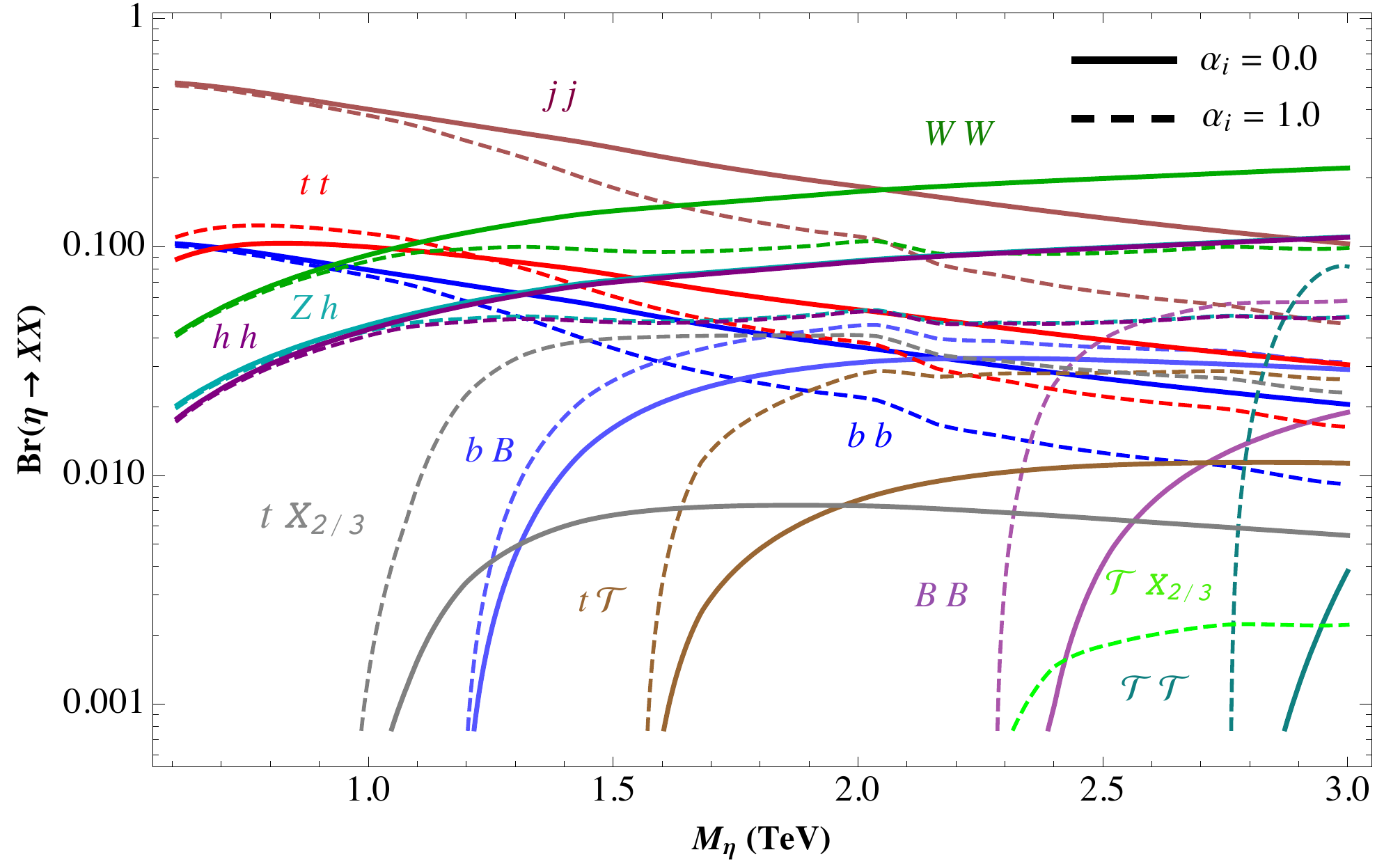}
\caption{\sf All branching ratios for the different $\eta$-decay modes at $\fA$ by setting $\xi=0.2$ and $a_\eta=1/2$, and accounting for no currents, \ie $\alpha_i=0$, as well as their contribution when $\alpha_i=1$ (thick and dashed curves).}
\label{Branching-ratios}
\end{center}
\end{figure}


Posterior decays of the heavy resonance  may occur into single and double top partner's  final states (the former accompanied by an associated SM quark), as well as into gauge and Higgs channels, \eg $\eta\to \{hh, WW, Zh\}$. The fermionic decay channels, for both of the single-double top partner's final states, will be triggered by the effective terms

\be
\LL_{Xf\eta}= \sum_{f=u,d}\left[g_{X f\eta}\,{\bar X}\,f\,\eta\,\,+\,\,{\bar X}\,\slashed{\partial}\eta\left(g_{X_L f_L \eta}\,P_L+ g_{X_R f_R \eta}\,P_R\right)f\right] \,\,+\,\,\hc,
\label{Partners-fermion-eta}
\ee

\be
\LL_{XX\eta}=g_{X X \eta}\,{\bar X}\,X\,\eta\,\,+\,\,{\bar X}\,\slashed{\partial}\eta\left(g_{X_L X_L \eta}\,P_L+ g_{X_R X_R \eta}\,P_R\right)X
\label{Partners-Partners-eta}
\ee

\nt with $X=\{T,B,\Xtt,\Xft,\Tt\}$. The latter couplings entail diagonalization effects from the elementary-composite sectors and are reported in Appendix~\ref{Effective-couplings}. Fig.~\ref{Branching-ratios} gathers the  branching ratios for two different cases $\alpha=0,1$ (thick-dashed curves) at $\fA$ with $\xi=0.2$  and by setting $a_\eta=1/2$ at~\eqref{eta-Lagrangian}, whose involved effective term is responsible for the gauge and Higgs channels.  Generically, these modes will be more relevant rather than the fermionic channels as a consequence of the involve kinematics of both initial and final states. Some comments are in order:

\begin{itemize}

\item  No fermion-resonance currents ($\alpha=0$) entails dominant dijet, top-pair and gauge, Higgs channels, while subdominant single-double partners final states. The dijet channel is the dominant one for $M_\eta\lesssim 2$ TeV, becoming subdominant with respect the $W$-pair for a higher mass value. 

\item The scenario is altered after switching extra fermion-resonance couplings on ($\alpha=1$). Indeed, the dijet, top-pair and gauge, Higgs channels qualitatively diminish, with a notorious enhancement for all the single and double partner final states in contrast. Despite this, the former modes are still relatively the dominant ones.

\item The enhancement occurring at the partner final states, may be a slight departure, as in the case of the mode $b\cB$, or even an rough increase of one or two orders of magnitude for the $t\Xtt$ and $\cT\cT$ channels.

\end{itemize}

\nt Analogous comments apply for the product of the scalar resonance production cross section times the corresponding branching ratio, not displayed here for brevity reasons. Once the scalar resonance are produced, their decays can generate, aside from the gauge and Higgs channels, either a single or double quark partner in the final states. A fuller top partner production mechanism is triggered by bringing  QCD, EW and Higgs-mediated interactions onto the stage. 

\section{Producing top partners and decays}
\label{Top-partner-production}
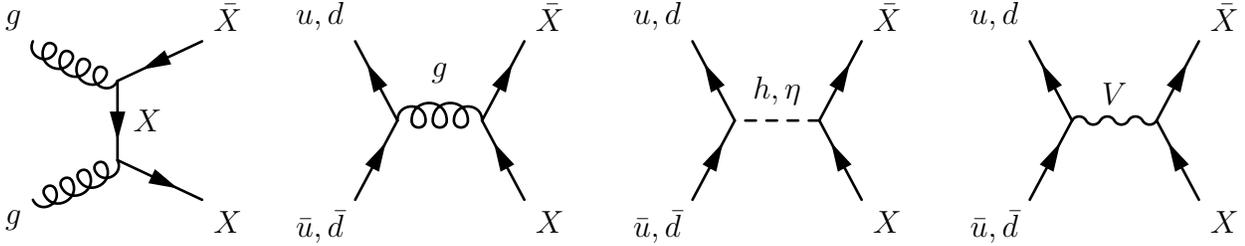
\begin{figure}
\hspace*{-0.7cm}
\begin{tabular}{cccc}
\input{Fdiagrams/DoubleProduction-1}&
\hspace*{0.5cm}
\input{Fdiagrams/DoubleProduction-3}&
\hspace*{0.7cm}
\input{Fdiagrams/DoubleProduction-2}&
\hspace*{0.7cm}
\input{Fdiagrams/DoubleProduction-4}
\vspace*{1cm}
\end{tabular}
\caption{\sf Diagrams contributing to the double partner production, where $V=Z,\gamma$ and with $X$ denoting any $X=\{T,B,\Xtt,\Xft,\Tt\}$. }
\label{Double-Production-diagrams}
\end{figure}

\nt All the quark partners are colored, hence their pair-production at hadron colliders is QCD-driven as it is shown in Fig.~\ref{Double-Production-diagrams}, being completely model-independent and insensitive to the degrees of compositeness of the associated SM quarks. Qualitatively, the top partner production is independent on whether both or only one multiplet is present in the effective theory.

\subsection{Double Partner production}
\label{Double-Partner-production}

\begin{figure}
\begin{center}
\hspace*{-0.5cm}
\vspace*{1cm}
\includegraphics[scale=0.42]{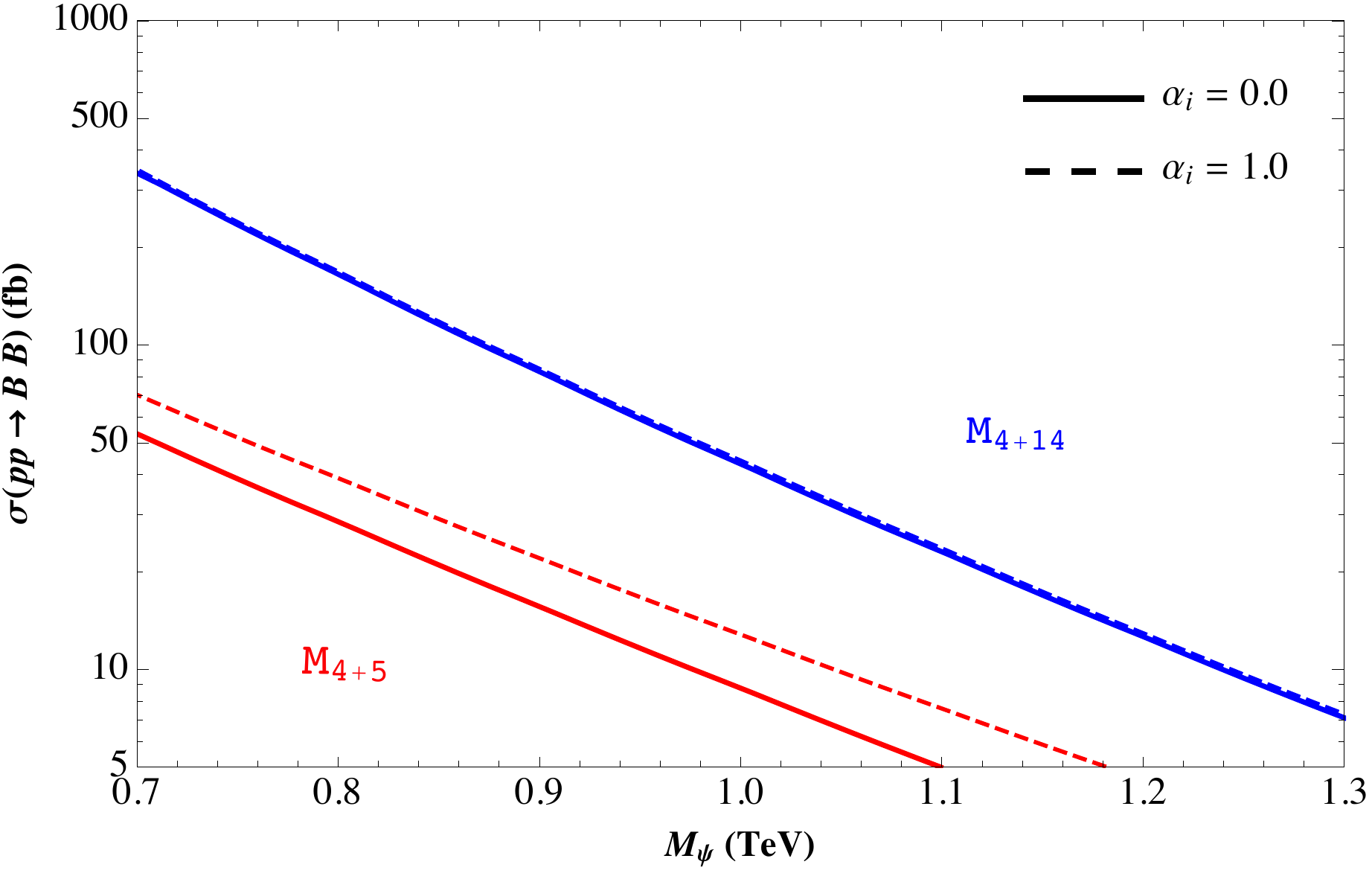}
\hspace*{0.7cm}
\includegraphics[scale=0.42]{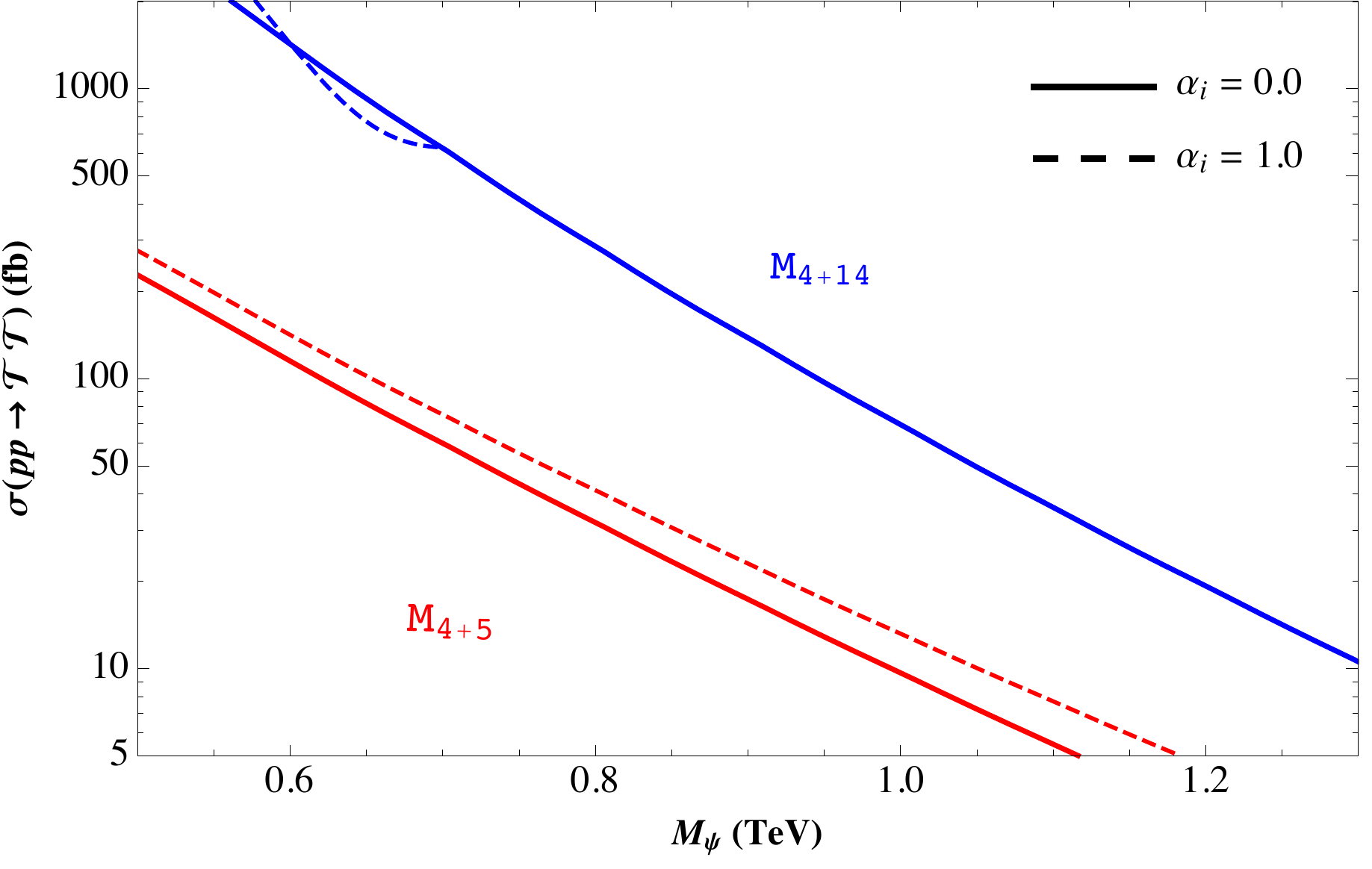}
\hspace*{-0.5cm}
\includegraphics[scale=0.42]{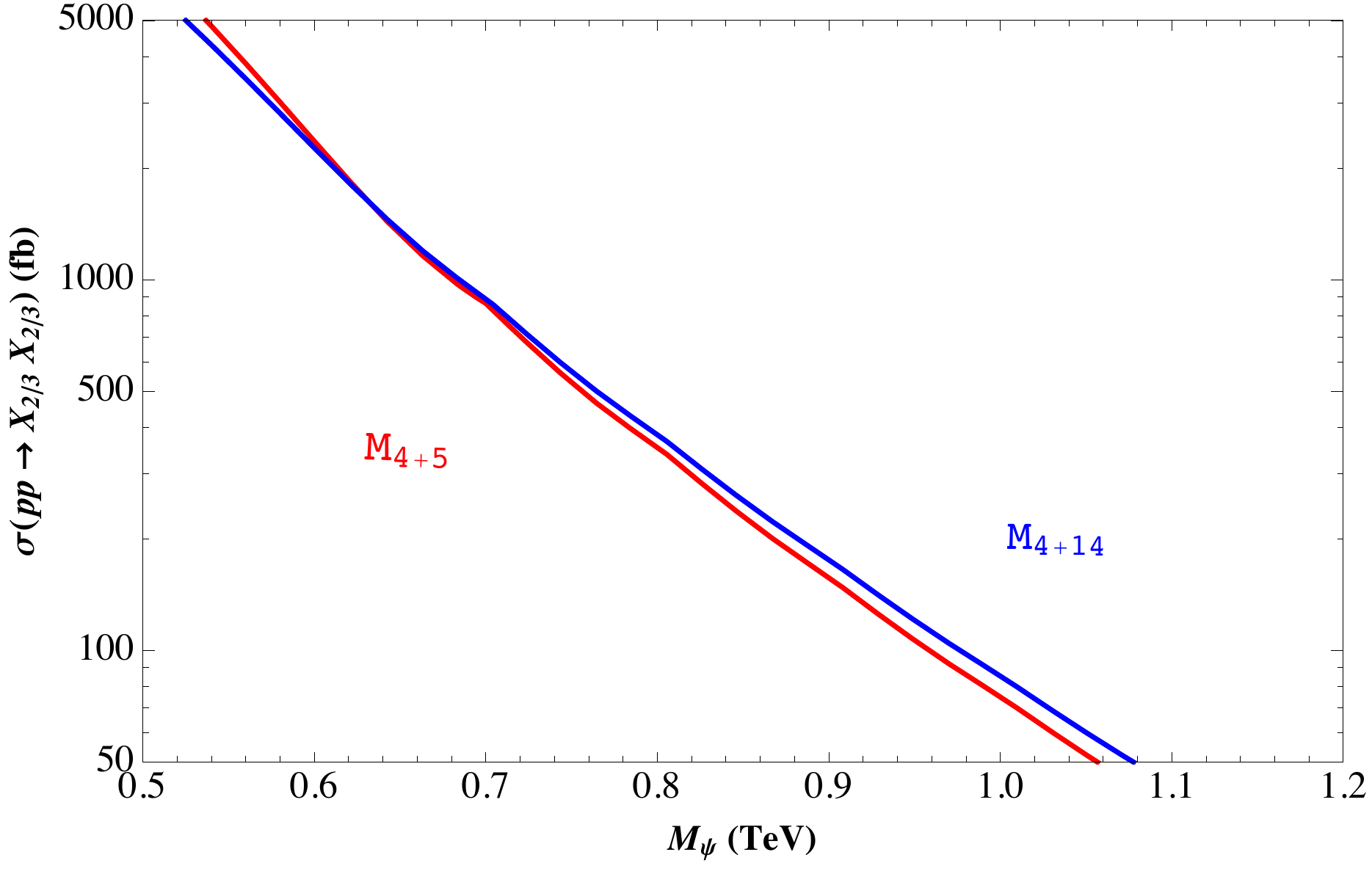}
\hspace*{0.7cm}
\includegraphics[scale=0.43]{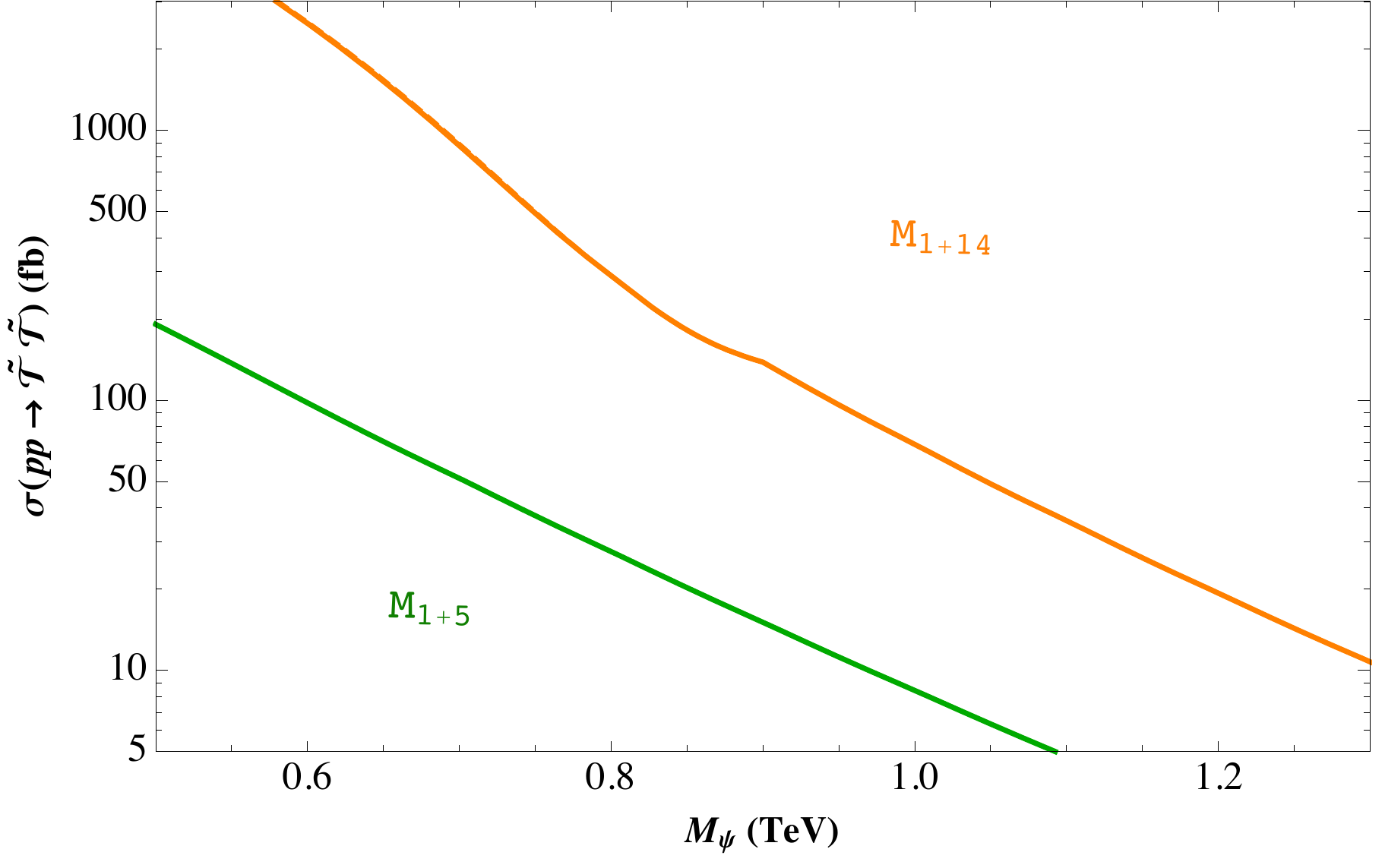}
\caption{\sf Double-partner production cross sections at 14 TeV  for $\xi=0.2$, only for neutral final states. Two different situations $\alpha=0,1$ (thick-dashed curves) are plotted to compare the impact on the production from the fermion-resonance Lagrangian $\LL_{\bf M\,+\,\eta}$ of~\eqref{Derivative-couplings}.}
\label{Double-partner-production}
\end{center}
\end{figure}

\nt The production of double-partner final states receives contributions from QCD as well as SM gauge, Higgs, and $\eta$-mediated processes as it is depicted in Fig.~\ref{Double-Production-diagrams}.  Such production is controlled by the model-dependent couplings $g_{ff \eta},\,g_{f_L f_L \eta},\,g_{f_R f_R \eta},\,g_{XX\eta},\,g_{X_L X_L \eta},\,g_{X_R X_R\eta}$ through~\eqref{Scalar-Yukawa-derivative-couplings} and~\eqref{Partners-Partners-eta}, and by the analogous ones involving the gluon and SM neutral gauge fields. Non-zero parametric-dependent modifications are induced as soon as extra fermion-resonance effects are considered. Fig.~\ref{Double-partner-production} collects double-partner production cross sections only for neutral final states, where we have constructed the pair cross sections for each value of the mass parameter $M_{\bf{4}}=M_{\bf{1}}=M_\Psi$ by interpolation using MadGraph 5 simulations, at 14 TeV LHC in all the models for $\xi=0.2$, and for a fixed scalar mass $M_\eta\sim 1.25$ TeV. Two different situations $\alpha=0,1$ (thick-dashed curves) display the impact on the production from the additional fermion-resonance effects regarded here. Slight enhancements occurs at $\fA$, whereas vanishing-tiny contributions are induced at the rest of the models due to the implied $f$-suppressed derivative couplings of~\eqref{Derivative-couplings}.

The final states $\cT\cT$ and $\cB\cB$ are dominantly produced via $pp$ collisions in $\fB$ as the involved quark partner masses are smaller than the corresponding ones at $\fA$ (see \eqref{Masses-expanded-5}-\eqref{Masses-expanded-14}  and Fig.~\ref{4plet-singlet-masses}). The final state $\Xtt\Xtt$ does not distinguish the elementary embeddings representation as the involved partner masses are equal at both models. The same comments apply qualitatively and quantitatively for the channel $\Xft\Xft$ as the involved partner masses are degenerate with the corresponding one for $\Xtt$ (see Appendix~\ref{Physical-fermion-masses}). Generically, producing pairs either of $\Xtt$ or $\Xft$ will be kinematically favoured with respect to the double production of both $\cT$ and $\cB$, because their relatively higher masses. Likewise, the pair production of the singlet $\widetilde{\cT}$ (Fig.~\ref{Double-partner-production}) is favoured at $\oB$, as the involved masses result smaller at $\bf{14}$-elementary embeddings compared with the one at $\bf{5}$-scenario (see Fig.~\ref{4plet-singlet-masses}, right plot).

\subsection{Single Partner production}
\label{Single-Partner-production}

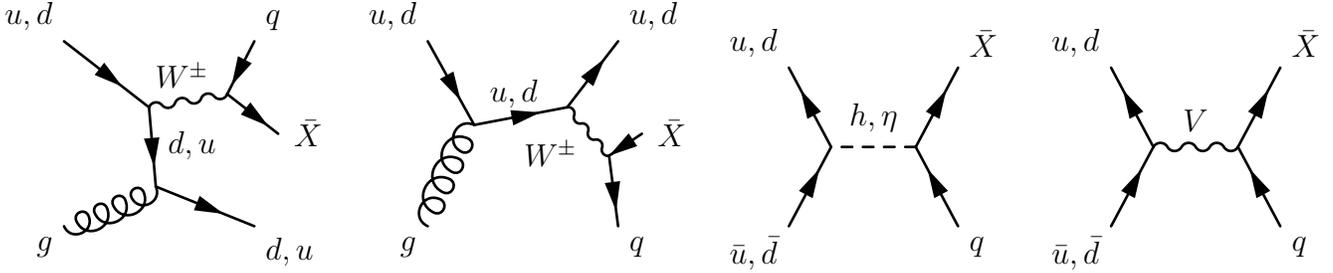
\begin{figure}
\hspace*{-0.8cm}
\begin{tabular}{cccc}
\input{Fdiagrams/SingleProduction-1}&
\hspace*{0.7cm}
\input{Fdiagrams/SingleProduction-3}&
\hspace*{0.7cm}
\input{Fdiagrams/SingleProduction-2}&
\hspace*{0.5cm}
\input{Fdiagrams/SingleProduction-4}
\vspace*{1cm}
\end{tabular}
\caption{\sf Diagrams contributing to the single partner production, where $V=Z,\gamma$ and with $q$ standing for any up down-like quark conveniently couple to $X=\{T,B,\Xtt,\Xft,\Tt\}$.}
\label{Single-Production-diagrams}
\end{figure}

\begin{figure}
\begin{center}
\hspace*{-0.6cm}
\includegraphics[scale=0.425]{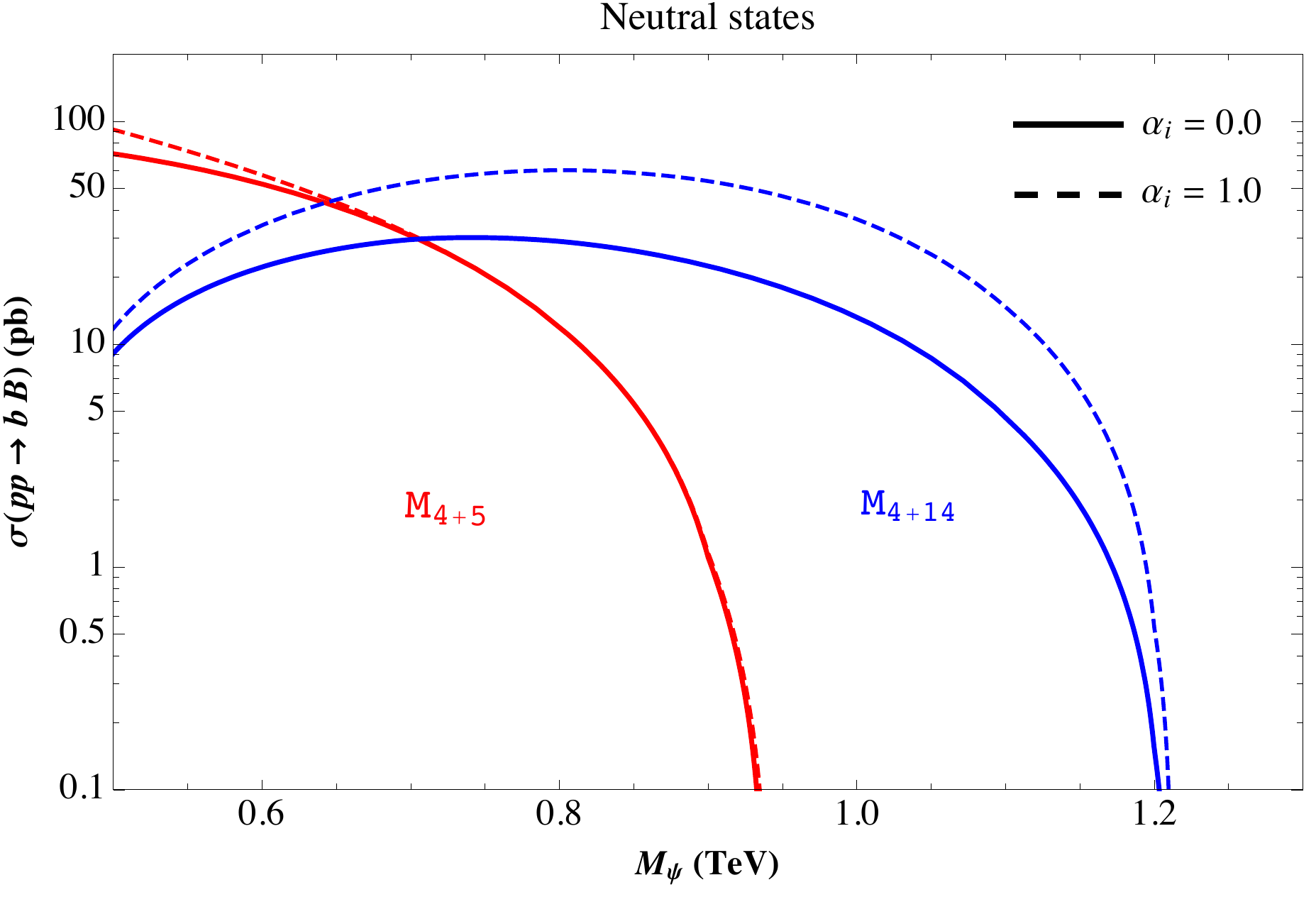}
\vspace*{0.3cm}
\hspace*{0.4cm}
\includegraphics[scale=0.4]{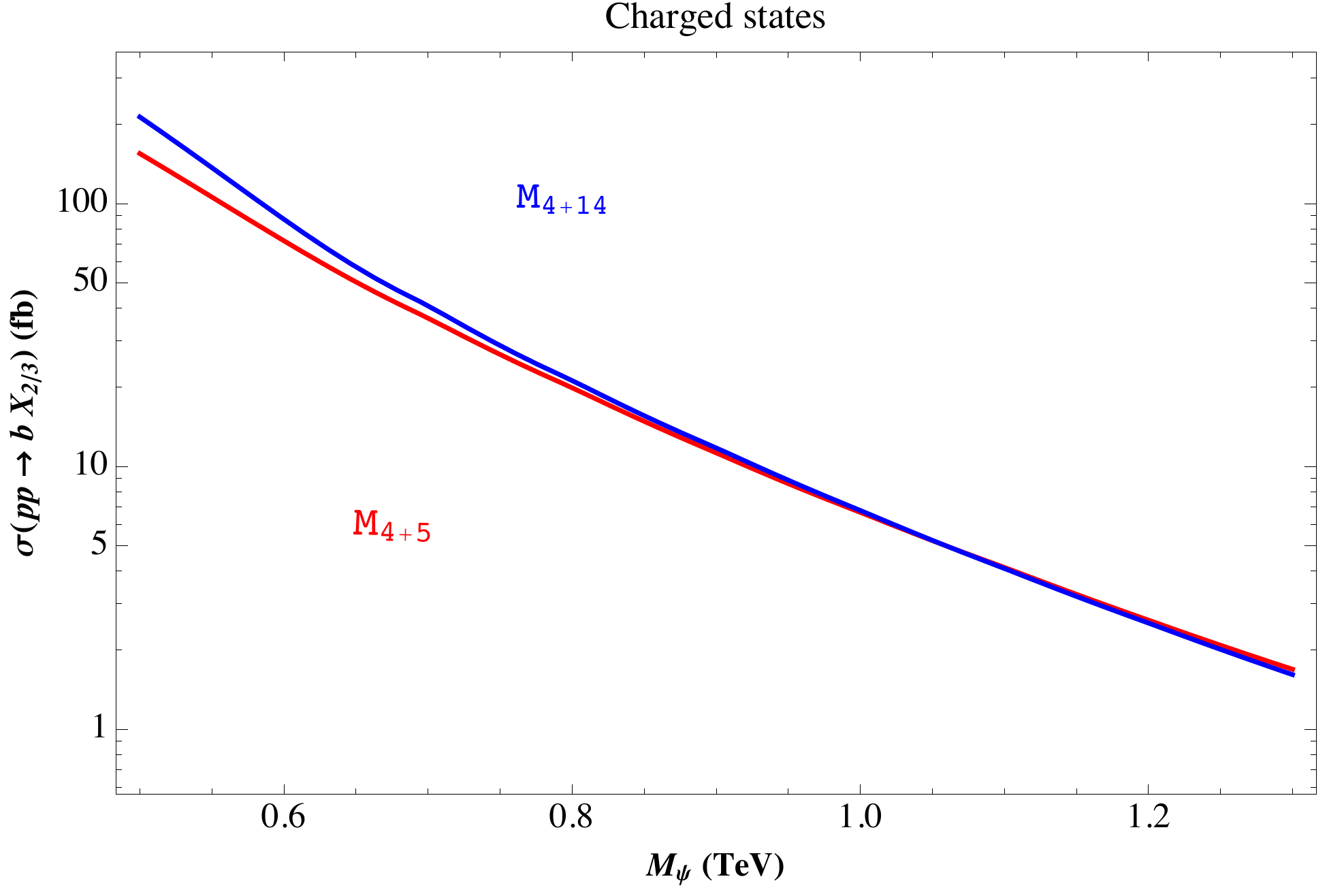}
\hspace*{-0.6cm}
\includegraphics[scale=0.435]{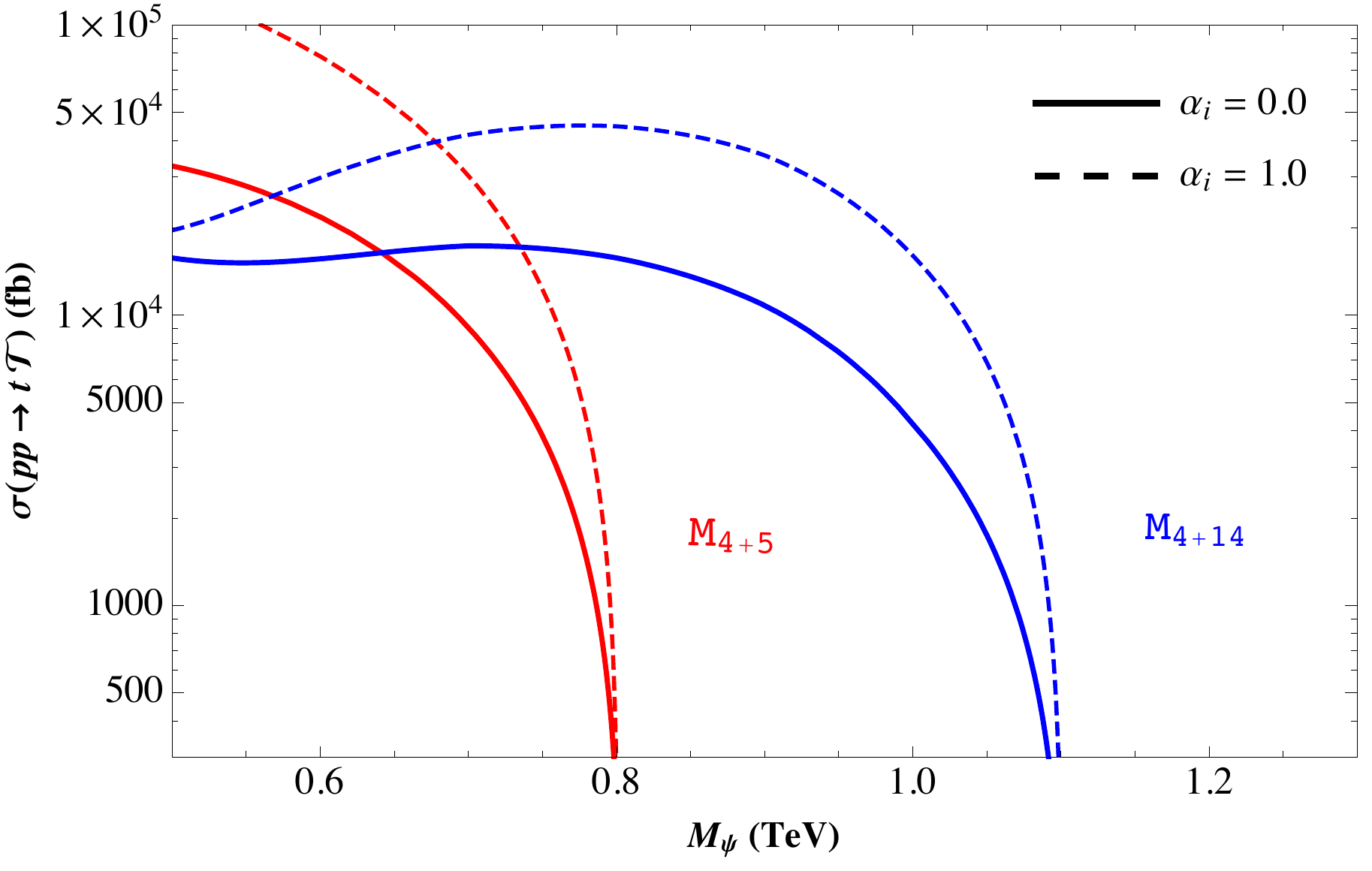}
\vspace*{0.3cm}
\hspace*{0.4cm}
\includegraphics[scale=0.395]{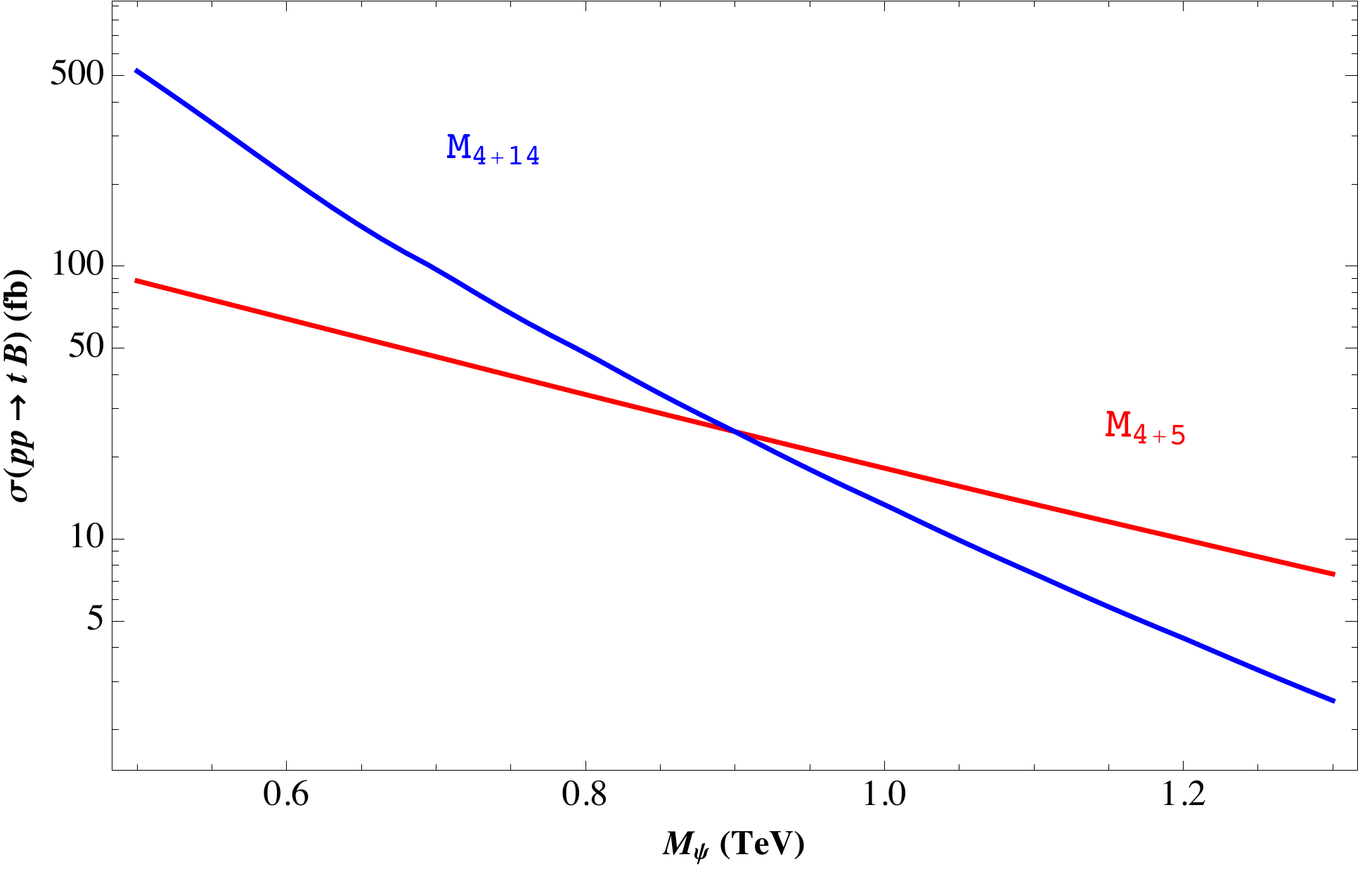}
\hspace*{-0.6cm}
\includegraphics[scale=0.42]{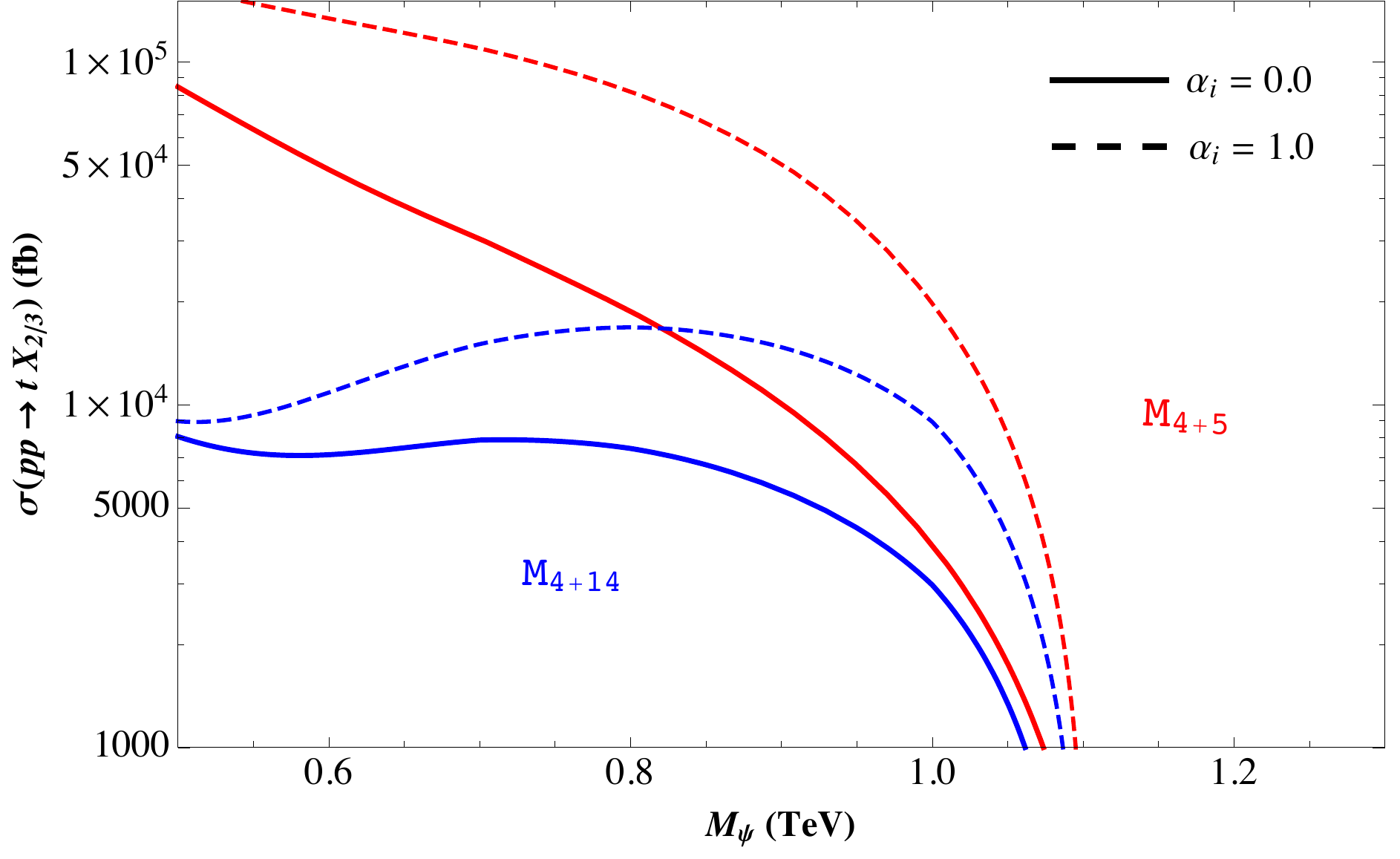}
\vspace*{0.3cm}
\hspace*{0.4cm}
\includegraphics[scale=0.395]{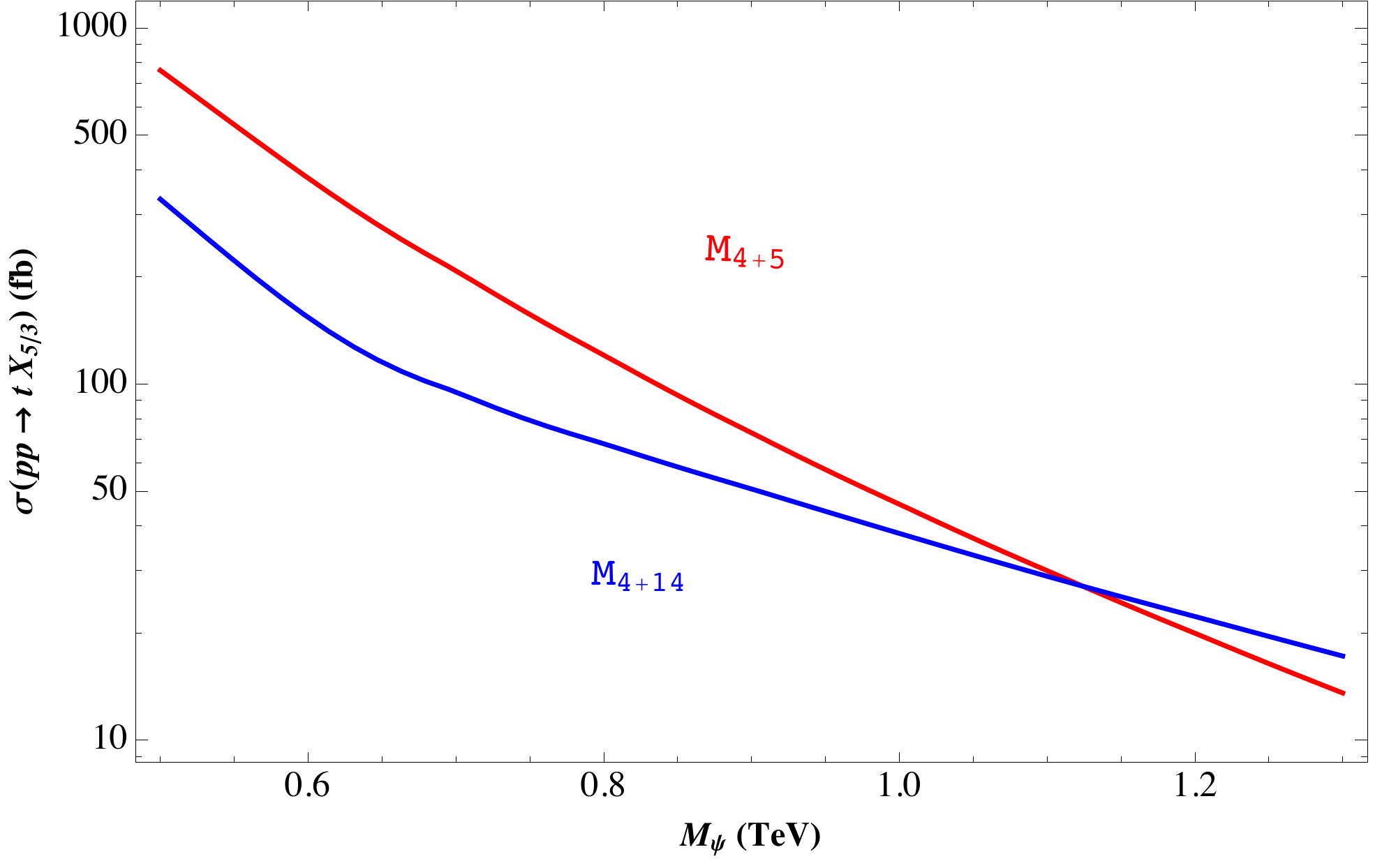}
\hspace*{-0.6cm}
\includegraphics[scale=0.44]{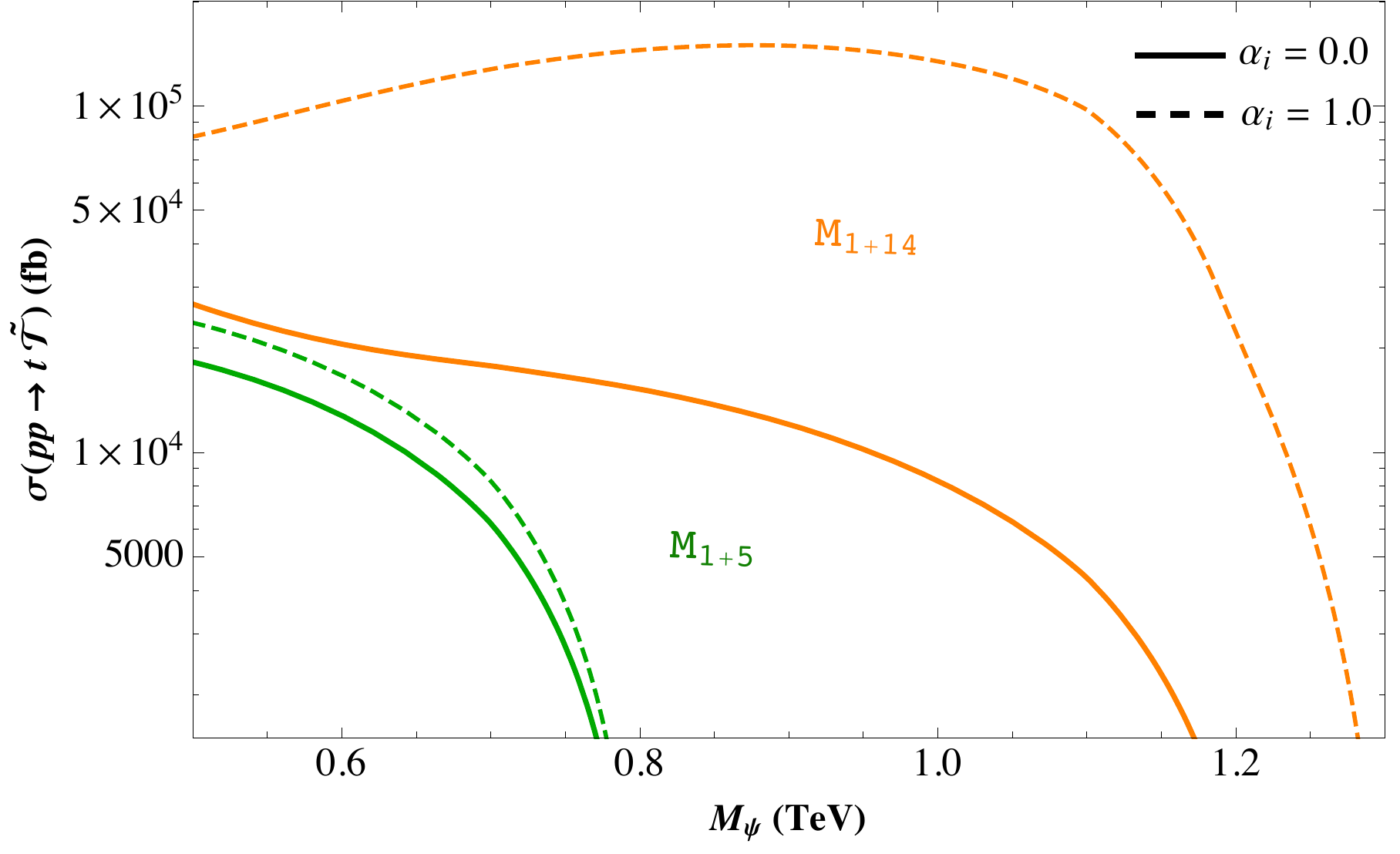}
\vspace*{0.3cm}
\hspace*{0.4cm}
\includegraphics[scale=0.4]{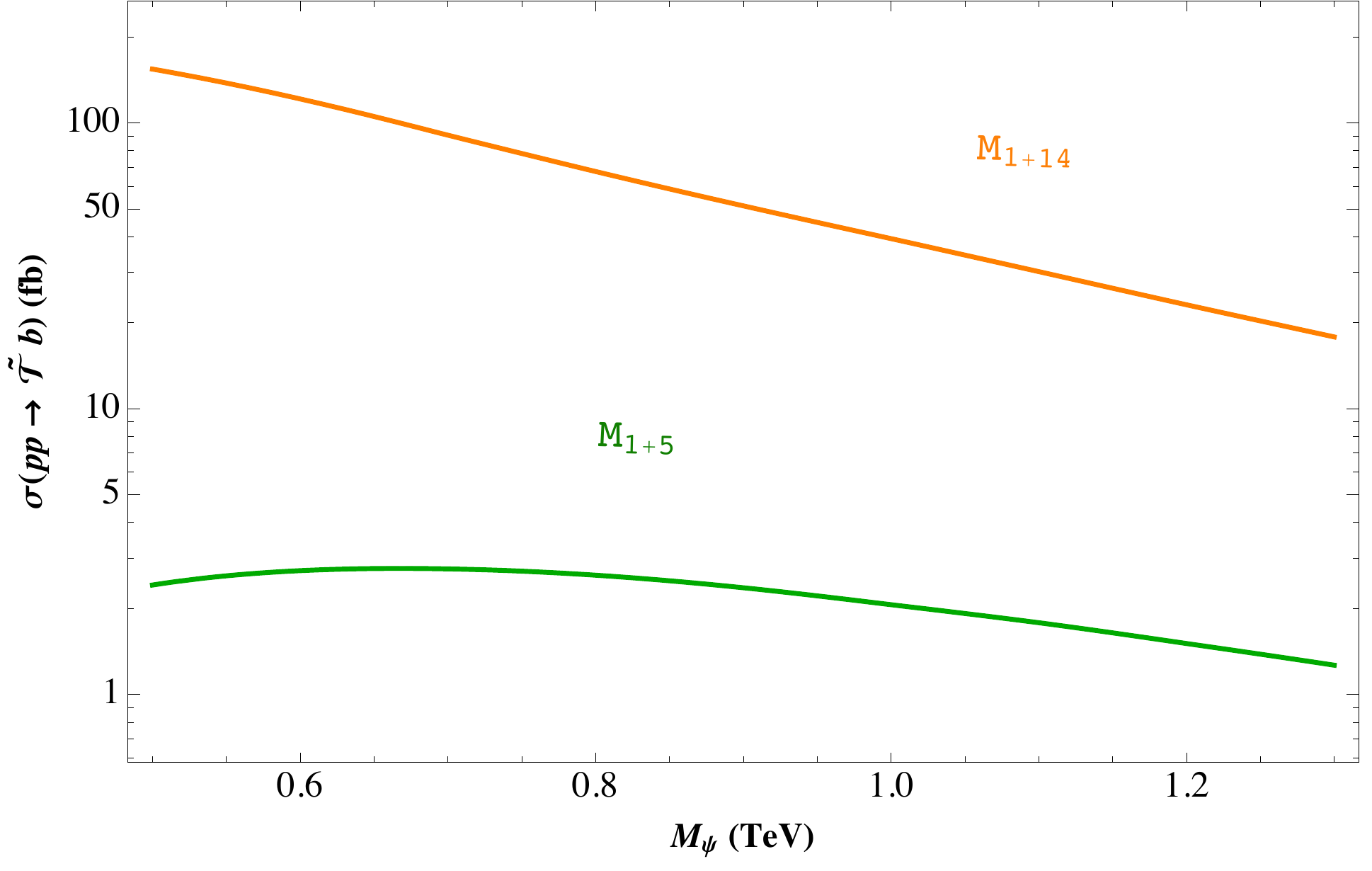}
\vspace*{-1.2cm}
\caption{\sf Single-partner production cross sections at 14 TeV  for $\xi=0.2$, either for the neutral final states (left) or the charged states (right). Two different situations $\alpha=0,1$ (thick-dashed curves) are plotted to compare the impact on the production from the fermion-resonance Lagrangian $\LL_{\bf M\,+\,\eta}$ of~\eqref{Derivative-couplings}. EW interactions are insensitve to the contributions of the currents, hence the charged final states are independent of their impact.}
\label{Single-partner-production}
\end{center}
\end{figure}

\nt QCD induces the production of single-partner final states, together with the SM gauge, Higgs and $\eta$-mediated processes for the case of neutral final states respectively, Fig.~\ref{Single-Production-diagrams}. These channels are gathered in Fig.~\ref{Single-partner-production}, where the mode $b\cT$ has been omitted for briefness reasons. Notice that the departures induced by the extra fermion-scalar couplings in~\eqref{Derivative-couplings} are only exhibited at the neutral final states as they are sensitive to the mediation of the scalar resonance $\eta$ via derivative couplings at $\LL_{\bf M\,+\,\eta}$. Cross section values are generically increased  by the presence of the latter couplings, becoming notoriously enhanced at the channels $t\Xtt$ and $t\widetilde{\cT}$ for the models $\fA$ and $\oB$ respectively in Fig.~\ref{Single-partner-production}, 3rd-4th plots left. The kinematic of less massive final states at the models $\fB$ and $\oB$ is responsible for the relative dominance of the former with respect $\fA$. The latter dominates compared with $\oA$ at the neutral channels $b\cB$, $t\cT$, $t\widetilde{\cT}$, and at the charged final states $b\Xtt$, $b\widetilde{\cT}$ as well. Although some cases do not obey this, like the mode $t\Xtt$ and $t\Xft$, where the combined effect of fermion diagonalization effects roughly suppress the induced contributions from the additional interactions of~\eqref{Derivative-couplings}. Despite the absence of the flavour-changing neutral couplings in the charge −1/3 sector~\cite{DeSimone:2012fs}, and of the $\cB \to hb$ channel at $\fA$, the final state $b\cB$ is still possible at the fourplet models via derivative couplings of~\eqref{Derivative-couplings} as it can be seen from Fig.~\ref{Single-partner-production}.

Yielding the singlet $\widetilde{\cT}$ at the $\bf{14}$-elementary embeddings results dominantly favoured rather than at the $\bf{5}$-scenario as the involved masses result smaller at the former model\footnote{Recently it has been analysed the top partner single productions through loops mediated by the scalar singlet in~\cite{Kim:2018mks}. With reasonable coupling strengths, the production rate of a top partner, in association with the SM top, can dominate top partner pair production at top partner masses higher than 1.5 TeV. See the reference for more details.}.  Likewise the production in pairs, the single productions will be controlled by the effective interacting terms among the fermions and the SM charged and neutral gauge, as well as by the model-dependent couplings $g_{ff \eta},\,g_{f_L f_L \eta},\,g_{f_R f_R \eta},\,g_{Xf\eta},\,g_{X_L f_L \eta},\,g_{X_R f_R\eta}$ along~\eqref{Scalar-Yukawa-derivative-couplings}-\eqref{Partners-fermion-eta}. These are computed analytically in our models, and they arise from 
the interactions reported in Appendices~\ref{Physical-fermion-masses} after performing the rotation to the physical basis of mass eigenstates. Explicit formulae for the couplings are straightforwardly derived and easily implemented in a {\sl{Mathematica}} notebook. 

Finally, some words concerning top partners decays are worth. The relevant couplings are encapsulated in~\eqref{Scalar-Yukawa-derivative-couplings}-\eqref{Partners-Partners-eta} and by the analogous ones involving SM charged and neutral gauge fields correspondingly. They can be computed analytically, and therefore precise tree-level formulae for the partial widths and for the branching fractions are obtained. Actually, more exotic channels are triggered, and if feasible, the kinematics would permit partners decaying into resonances in association with SM fermions or even in companion of partners less massive than the decaying initial state. Decays like $X\rightarrow X' V$ or $X'H$ arise in our models, and depending on the chosen parameters, they would either enhance or decrease some standard SM final states, and would strongly depend on the resonance mass spectrum as well as on the decaying partner mass. Exotic channels like $X\rightarrow X' \eta$ are theoretically allowed but less relevant though, as they involve higher masses in the final states\footnote{For a more detailed discussion on relevant decays see~\cite{DeSimone:2012fs} and  for a more recent update check~\cite{Chala:2017xgc,Bizot:2018tds,Banerjee:2017wmg}. Early discussions on the discovery potential of top-partners in a realistic composite Higgs model with LHC data can be found in~\cite{Dissertori:2010ug,Vignaroli:2012nf}.}. In a future work, we will explore these issues and the flexibility entailed by the parametric dependence for the feasibility of the SM and exotic decay channels of the involved partners.

The constraints on the top partners inferred from presently available LHC searches have been recently explored in~\cite{Chala:2017xgc,Bizot:2018tds}, imposing direct bounds on heavy top-like quarks with standard and exotic decays, meanwhile they have been implemented in~\cite{Yepes:2018dlw} in the context of top partners-vector resonances in CHMs. Similar constraints on the allowed parameter space of our models are obtained by the imposition of recent LHC partner searches.

\section{Parameter spaces and constraints} 
\label{Some-parameter-spaces}

\nt We have derived here the parameter spaces allowed by the recent available LHC partner searches, in terms of $\xi$ and the mass scales $M_\eta$ and $M_\Psi$. Recently, CMS has released~\cite{Sirunyan:2017pks} the results of searches for vector-like quarks, 2/3 and -4/3-electrically charged, that are pair produced in $pp$ interactions at $\sqrt{s} = 13$TeV, and decaying exclusively via the $Wb$ channel. Events were selected requiring a lepton and neutrino from one $W$, and a quark-antiquark pair from the other boson gauge. The selection requires a muon or electron, significant missing transverse momentum, and at least four jets. A kinematic fit assuming a pair production of 2/3 or -4/3 electrically charged vector-like quarks  was performed and for every event a corresponding candidate quark mass was reconstructed.  Upper limits were set in~\cite{Sirunyan:2017pks} for the pair production cross sections as a function of the implied vector-like quark masses. By comparing these limits with the predicted theoretical cross section of the pair production, the production of 2/3  or -4/3 electrically charged vector-like quarks is excluded at 95\% confidence level for masses below 1295 GeV (1275 GeV expected). More generally, the results set upper limits on the product of the production cross section and branching fraction to $Wb$ for any new heavy quark decaying to this channel. Such limits have been imposed in $\sigma \times Br$ for all of our models and are translated into exclusion regions for the parameter spaces involved by $\xi$, $M_\eta$ and $M_\Psi$. Computation of $Br\left(\cT\to Wb\right)$ and $Br(\widetilde{\cT}\to Wb)$ is performed including a scalar resonance in the final states for the total width, with a posterior simulation via MadGraph 5 of the pair production cross section of $\cT\cT$ and $\widetilde{\cT}\widetilde{\cT}$ at $\sqrt{s} = 13$ TeV for the fourplet and singlet models respectively. Fig.~\ref{MPsi-Xi} gathers the allowed parameter spaces $\left(M_\Psi,\,\xi\right)$ for all the fourplet and singlet models, with a total decay width summing the standard modes $W b$, $Z t$ and  $h t$ up (1st-2nd plots), and augmented by $\eta\,t$ (3rd-4th graphs). Consequently, the branching ratio for any channel will be also $M_\eta$-dependent and will entail a parametric dependence on the extra fermion-resonance interactions regarded here in~\eqref{Derivative-couplings}. Their impact is scanned along two different situations: the dashed border regions stand for the allowed parameter spaces assuming extra fermion-resonance couplings weighted by $\alpha=1$, whilst the others zones denote no additional interactions, \ie $\alpha=0$. The scalar resonance mass is fixed at the benchmark value $M_\eta=3$ TeV at the 3rd-4th graphs. Some comments are in order:
\begin{figure}
\begin{center}
\includegraphics[scale=0.199]{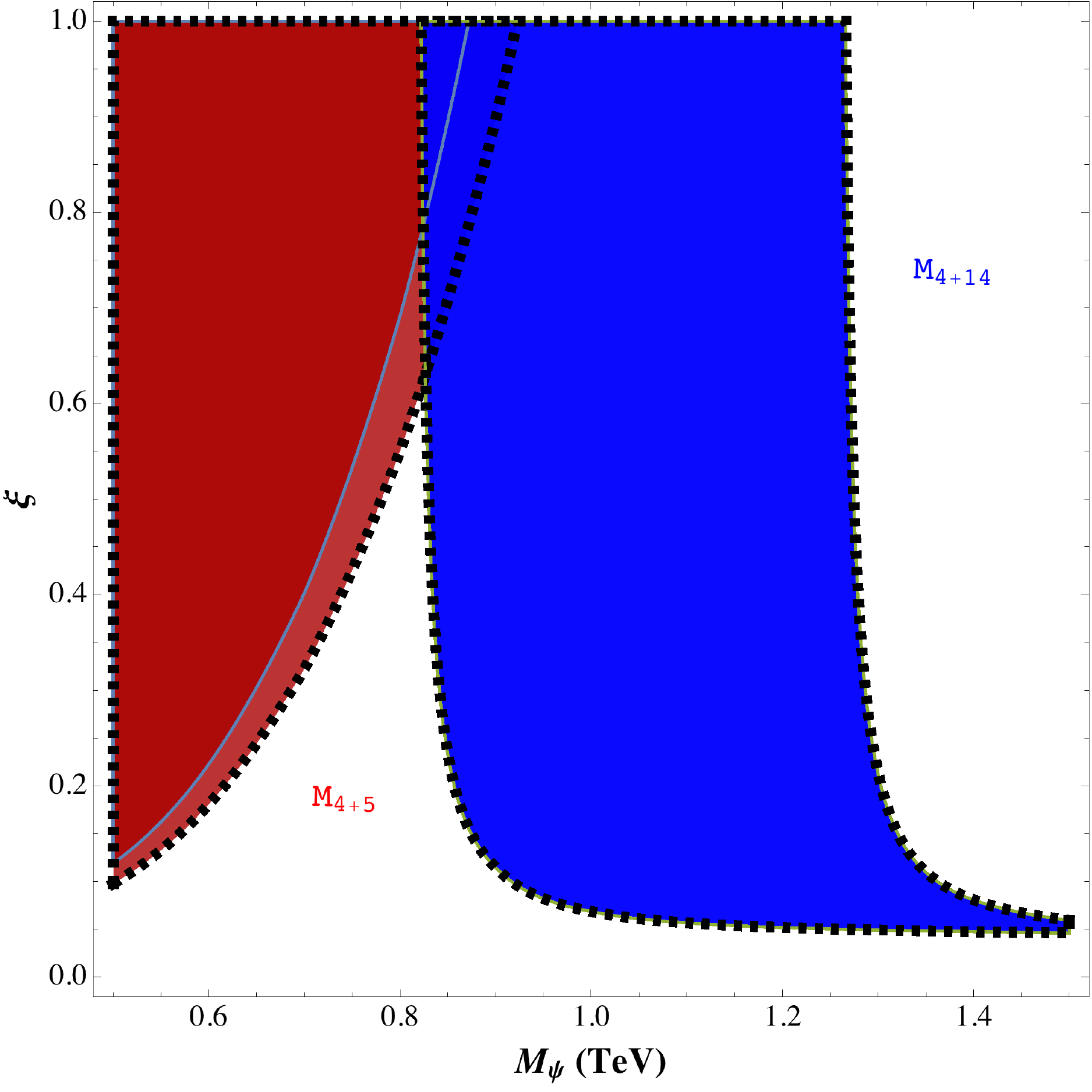}
\hspace*{0.12cm}
\includegraphics[scale=0.193]{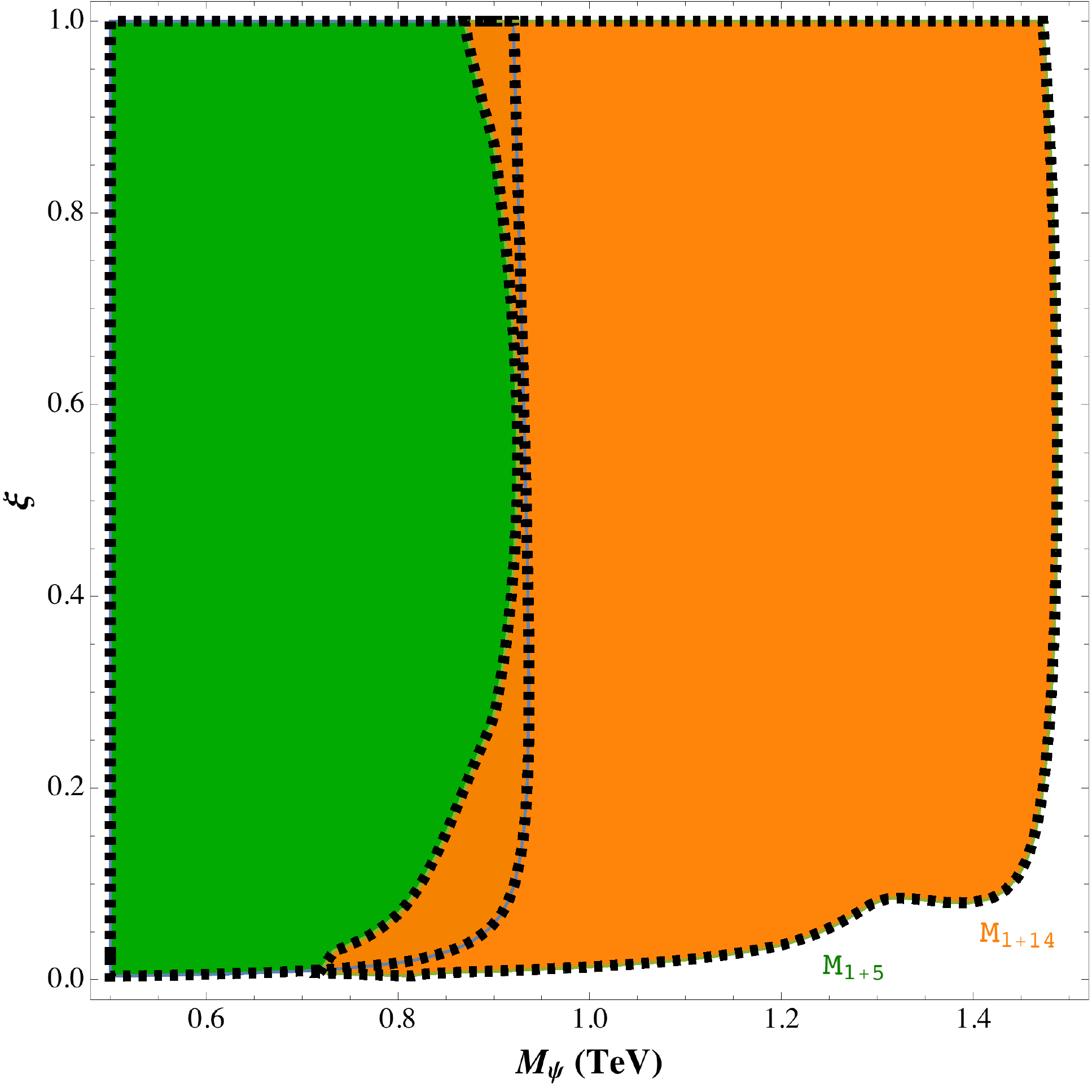}
\hspace*{0.12cm}
\includegraphics[scale=0.195]{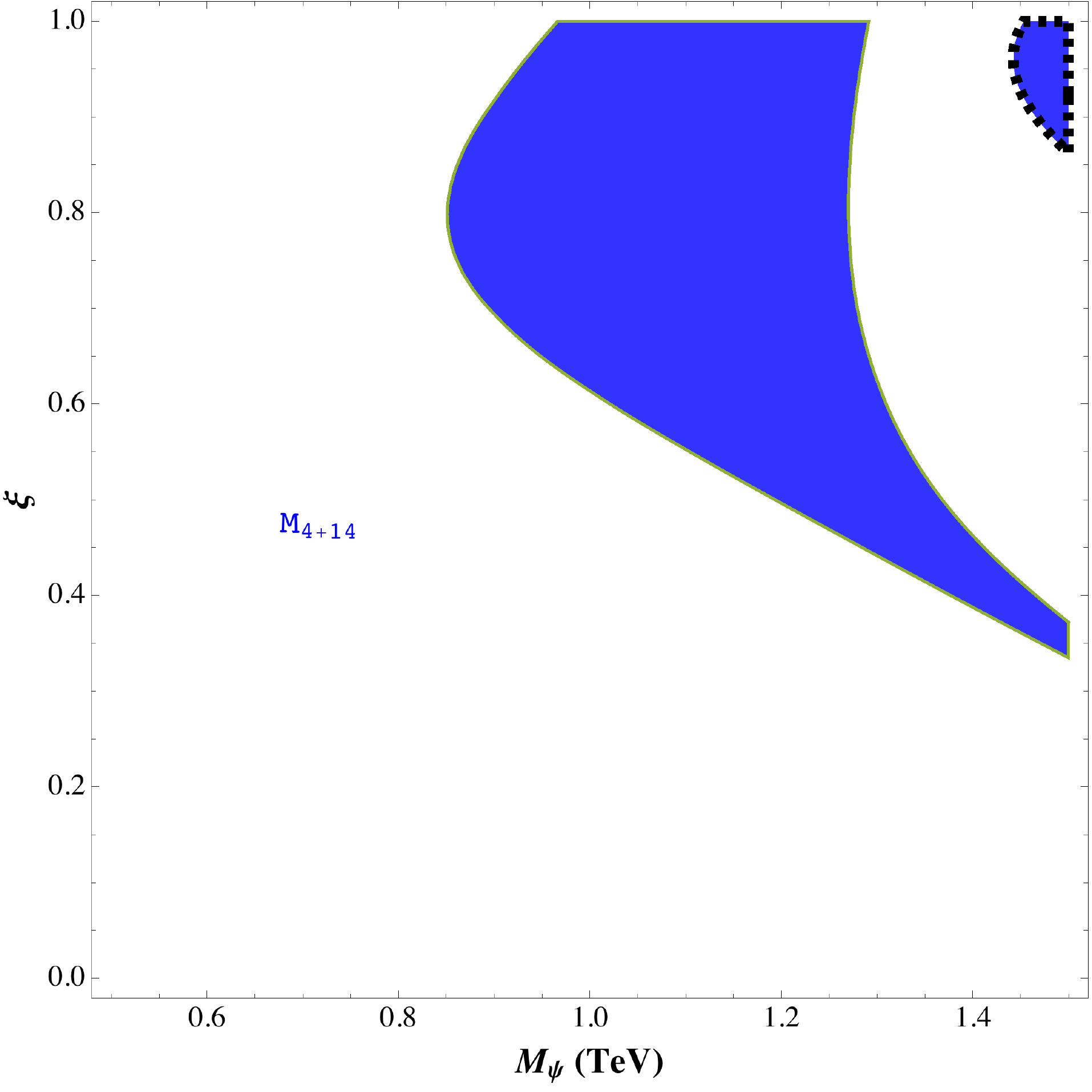}
\hspace*{0.12cm}
\includegraphics[scale=0.195]{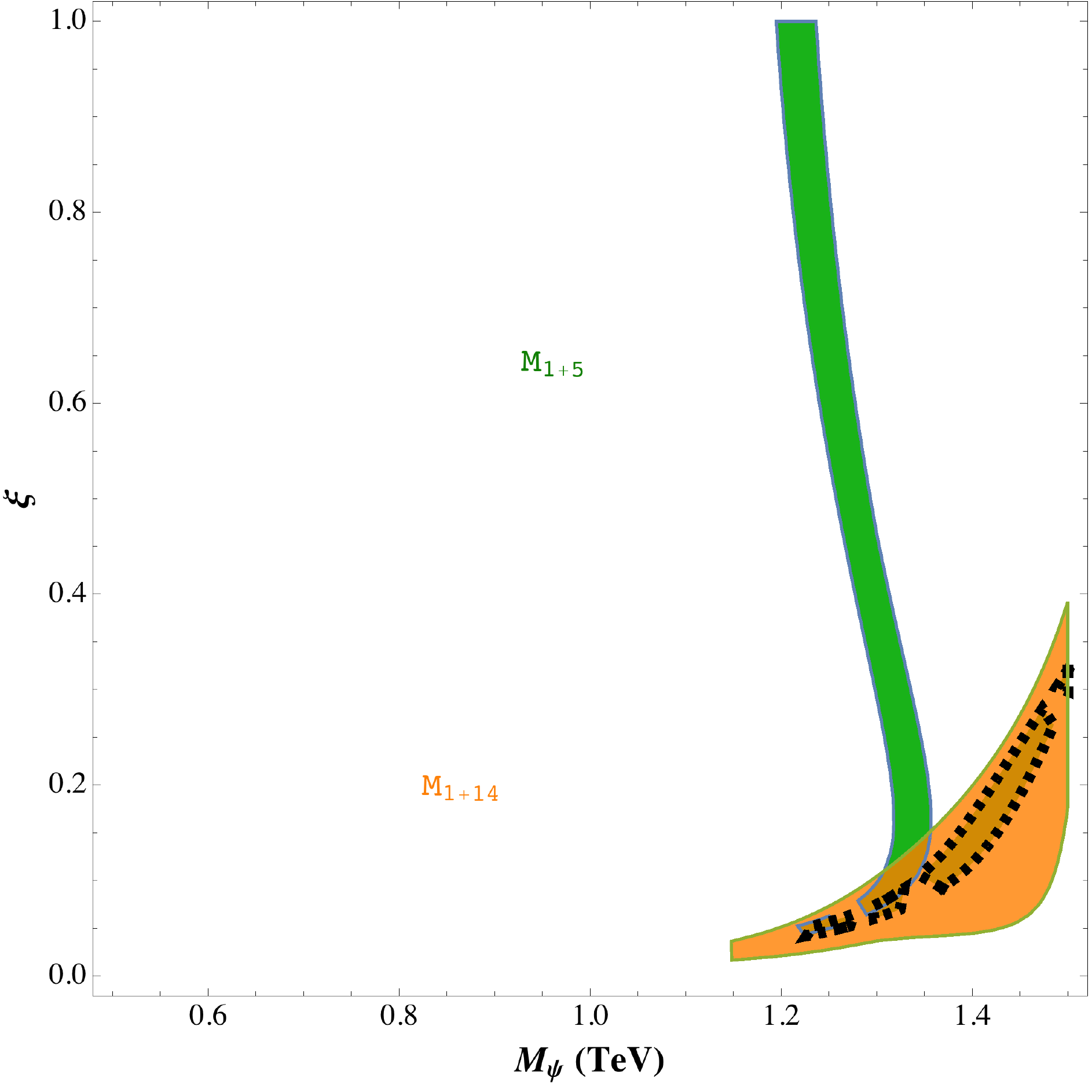}
\caption{\sf Parameter space $\left(M_\Psi,\,\xi\right)$ obtained from recent bounds on top partner searches through top-like decays into $W b$ final states prior to the inclusion of the scalar $\eta$ into the final states (1st-2nd plots) and after its inclusion by setting $M_\eta=3$ TeV (3rd-4th graphs). Recent bounds~\cite{Sirunyan:2017pks} on top partner searches through top-like decays into $W b$ final states have been imposed at all models. Two situations have been explored $\alpha=0,1$ (thick-dashed border).}
\label{MPsi-Xi}
\end{center}
\end{figure}

\begin{itemize}

\item If the scalar resonance $\eta$ is included into the final states, the allowed region is strongly constrained (3rd-4th plots), becoming further restricted to the tiny regions
\begin{align*}
\fB:&\quad\xi\sim [0.85,\,1.0]\quad \text{for}\quad M_\Psi\sim [1.45,\,1.5]\,\text{TeV}\quad\Rightarrow\quad m_\cT\sim [1.57,\,1.64]\,\text{TeV}\\
\oA:&\quad\xi\sim [0.05,\,0.3]\quad \text{for}\quad M_\Psi\sim [1.2,\,1.5]\,\text{TeV}\quad\,\,\,\Rightarrow\quad m_{\widetilde{\cT}}\sim [0.96,\,1.4]\,\text{TeV}
\end{align*}

if the extra couplings in~\eqref{Derivative-couplings} are included. $\fA$ turns out to be excluded at this point. The latter mass ranges are totally allowed at $\fB$ by the recent limits~\cite{Sirunyan:2017pks} on the exclusion at 95\% confidence level for masses below 1295 GeV, while partly permitted at $\oA$. Furthermore, the EWPT bounds, the vector resonance direct production bounds at LHC, as well as the expected LHC single Higgs production, and the double Higgs production at CLIC (see discussion in~\ref{Interplay}) are compatible with the $\xi$-range at both models. 

\item Disregarding the scalar resonance at the final states, a broader parameter space is allowed and the previous ranges become relaxed (1st-2nd plots). Indeed, $\xi$-intervals compatible with experimental expectations exist at both fourplet models, becoming ruled out at $\fA$ as they wholly fall inside the exclusion limit of~\cite{Sirunyan:2017pks}. At $\fB$, such bounds entail (see Fig.~\ref{4plet-singlet-masses})
\be
\fB:\quad 0.05\lesssim\xi\lesssim 0.3\quad \text{for}\quad M_\Psi\gtrsim 1.1\,\text{TeV}
\ee
favouring a extreme part of the obtained parameter space in Fig.~\ref{MPsi-Xi}. Likewise, the exclusion limit in~\cite{Sirunyan:2017pks} leads to
\begin{align*}
\oA:&\quad 0.01\lesssim\xi\lesssim 0.1\quad\,\,\, \text{for}\quad M_\Psi\sim [0.5,\,0.825]\,\text{TeV},\\
\oB:&\qquad\qquad \xi\gtrsim 0.15\quad\,\text{for}\quad M_\Psi\sim [1.35,\,1.48]\,\text{TeV}
\end{align*}
approving a small region for the associated parameter spaces at both models. Small $\xi$-values are allowed, being compatible with experimental constraints at both singlet models. Similar comments apply when switching the extra couplings in~\eqref{Derivative-couplings}, as their effect do not alter the implied parameter spaces.

\end{itemize}

\begin{figure}
\begin{center}
\includegraphics[scale=0.31]{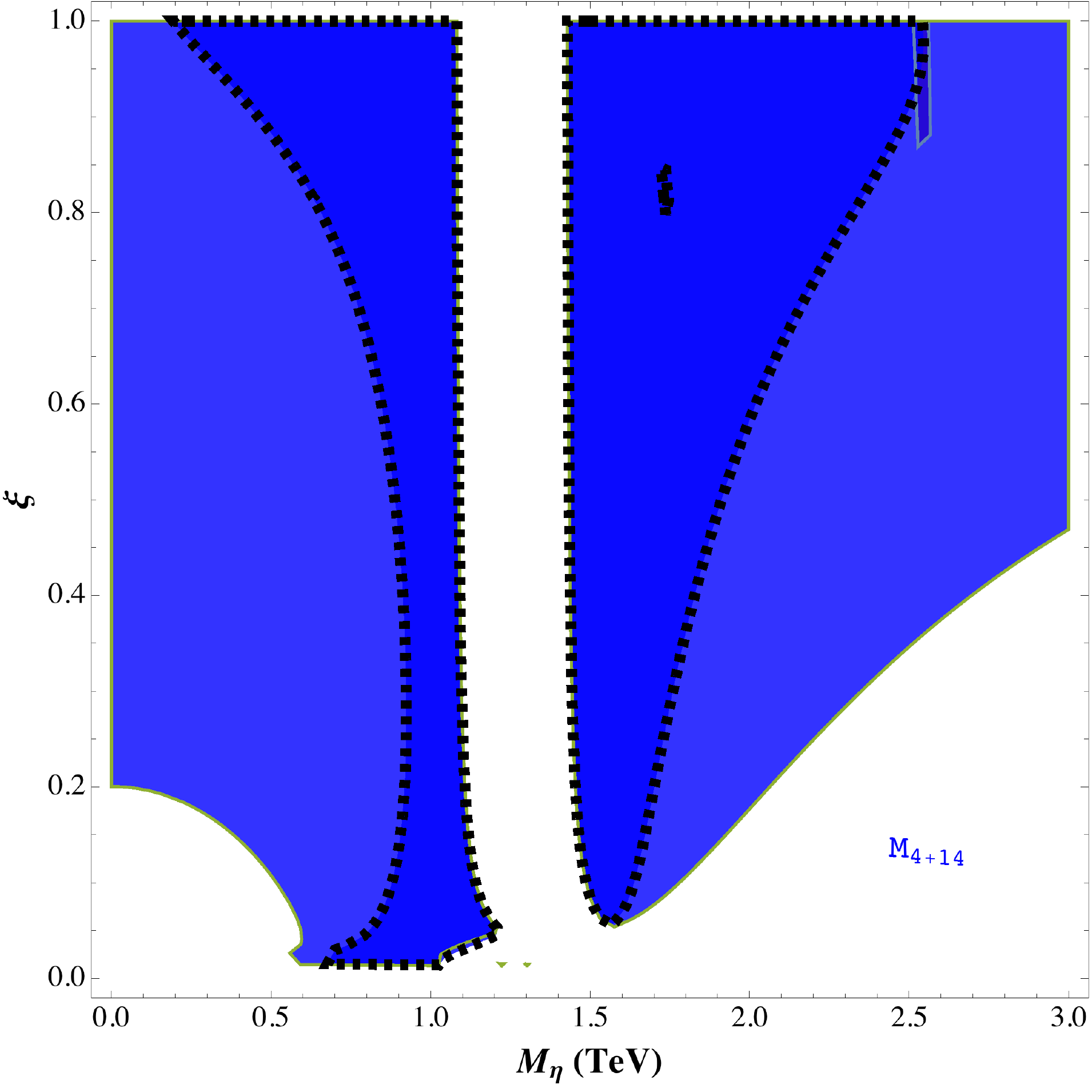}
\hspace*{1.5cm}
\includegraphics[scale=0.31]{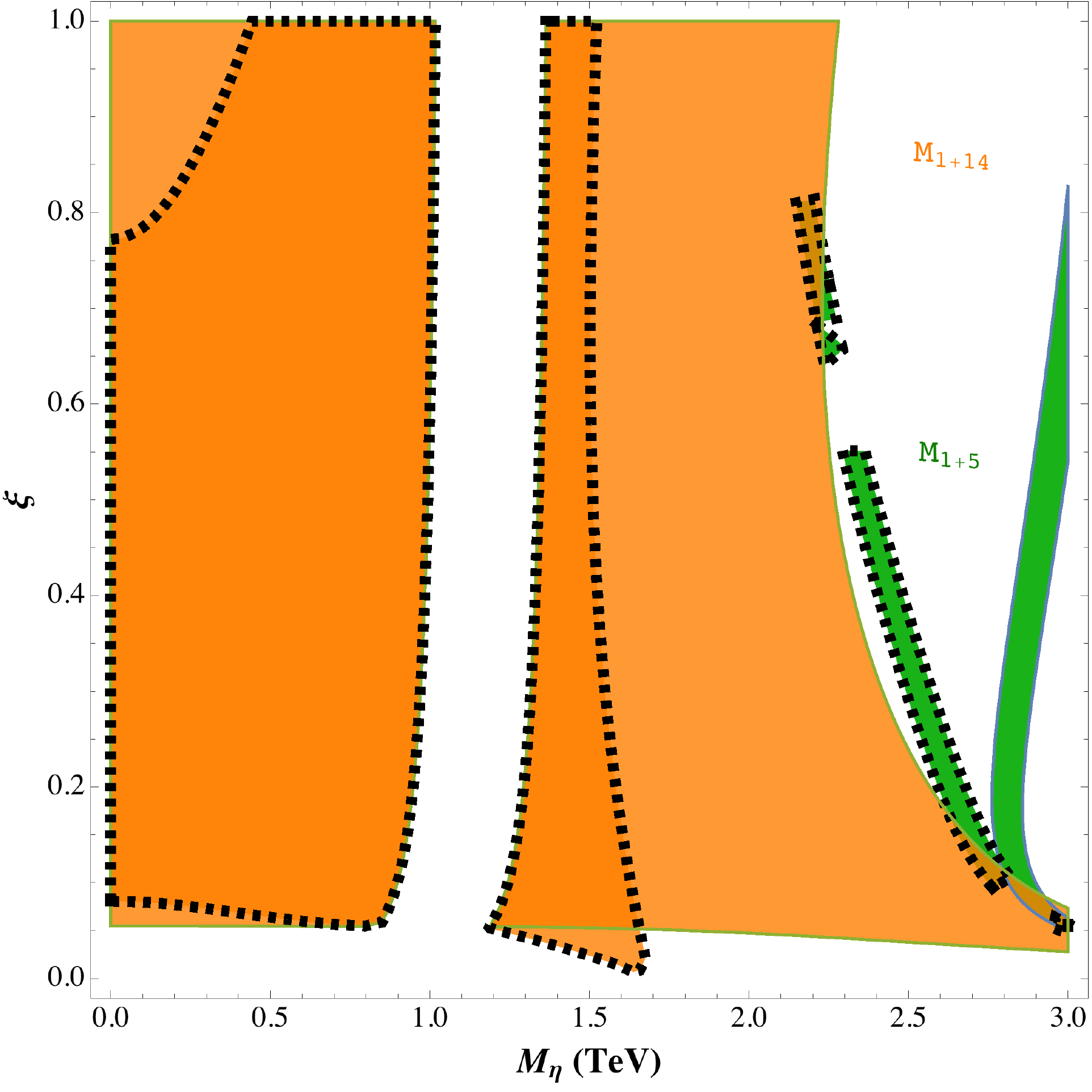}
\caption{\sf Parameter space $\left(M_\eta,\,\xi\right)$ by fixing $M_\Psi=1250$ GeV. Two situations have been explored, $\alpha=0,1$ (thick-dashed border).}
\label{Meta-Xi}
\end{center}
\end{figure}

\nt As a conclusion, the recent upper limits on top-like partners production permit part of the parameter spaces from $\fB$ and from the singlet models if the scalar resonance $\eta$ is disregarded, and whether the extra fermion-scalar interactions are considered or not. By including the scalar field into the final states a strongly bounded region, further constrained if the extra interactions in~\eqref{Derivative-couplings} are included, remains at $\fB$ and $\oA$ only. In this sense, those extra couplings are useful in discerning models and refining further their involved parameter space. An additional insight into the parametric freedom of the assumed scenarios can be explored by fixing the partner mass scale and letting the scalar resonance one to vary. This entails of course the scalar resonance inclusion at the final states. Fig.~\ref{Meta-Xi} illustrates this by setting $M_\Psi=1.25$ TeV\footnote{A bit below the threshold for the exclusion limit~\cite{Sirunyan:2017pks}.}. The parameter spaces are notoriously split into a left and right-handed regions, with the intermediate excluded ranges $M_\eta\sim$\, 1.2-1.4\,TeV and $M_\eta\sim$\, 0.9-1.2\,TeV  at the $\bf{14}$-embedding scenarios, and with $\oA$ favouring the higher scalar mass range $\sim$\, 2.8-3\,TeV. The inclusion of extra derivative couplings reduce a bit the higher and lower range masses at $\fB$, shifting the range masses to a lower one at $\oA$, while strongly constraining the right handed region to $M_\eta\sim$\, 1.2-1.6\,TeV at $\oB$. Consistent $\xi$-values are still feasible.
\begin{figure}
\begin{center}
\includegraphics[scale=0.326]{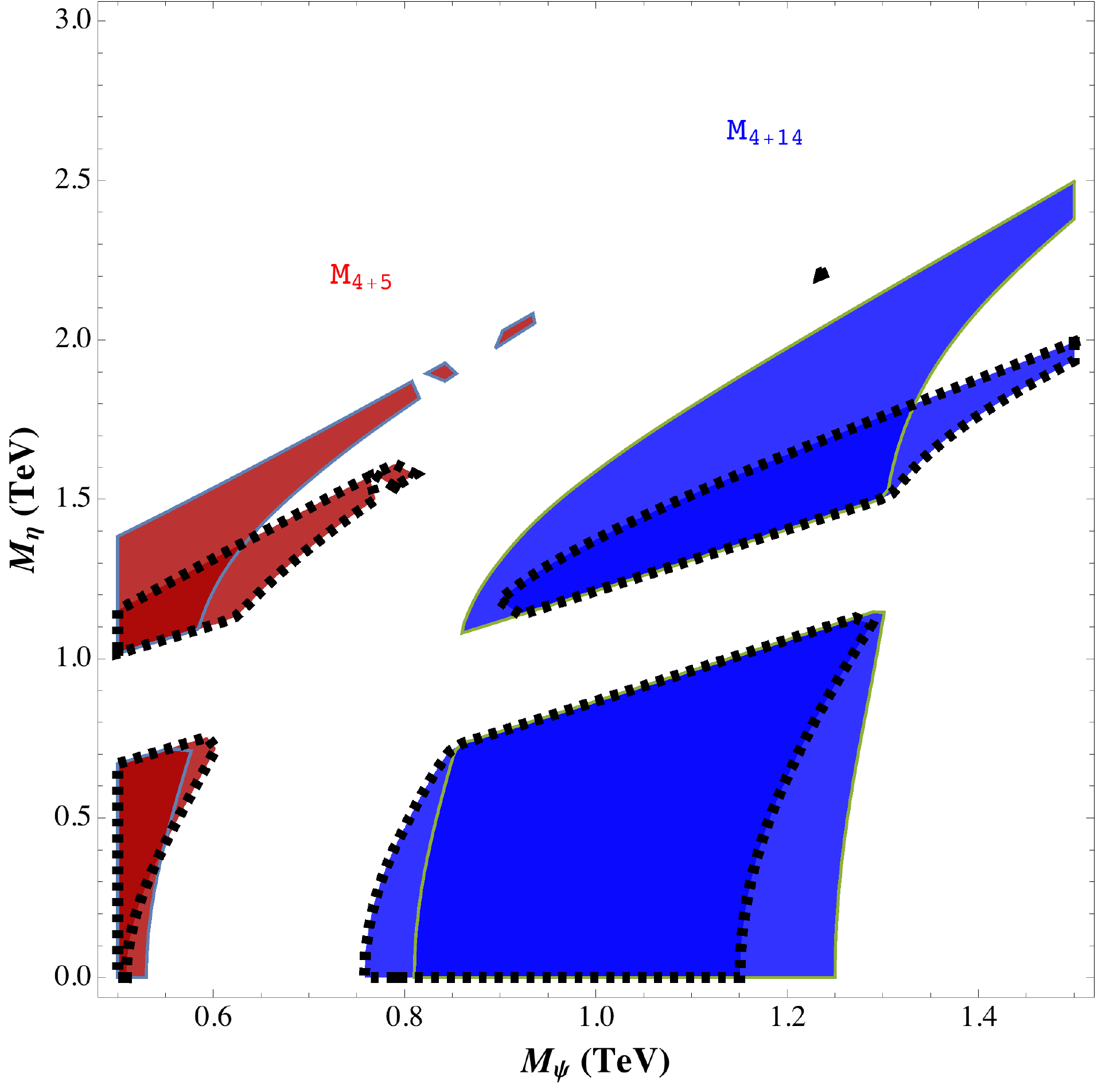}
\hspace*{1.5cm}
\includegraphics[scale=0.325]{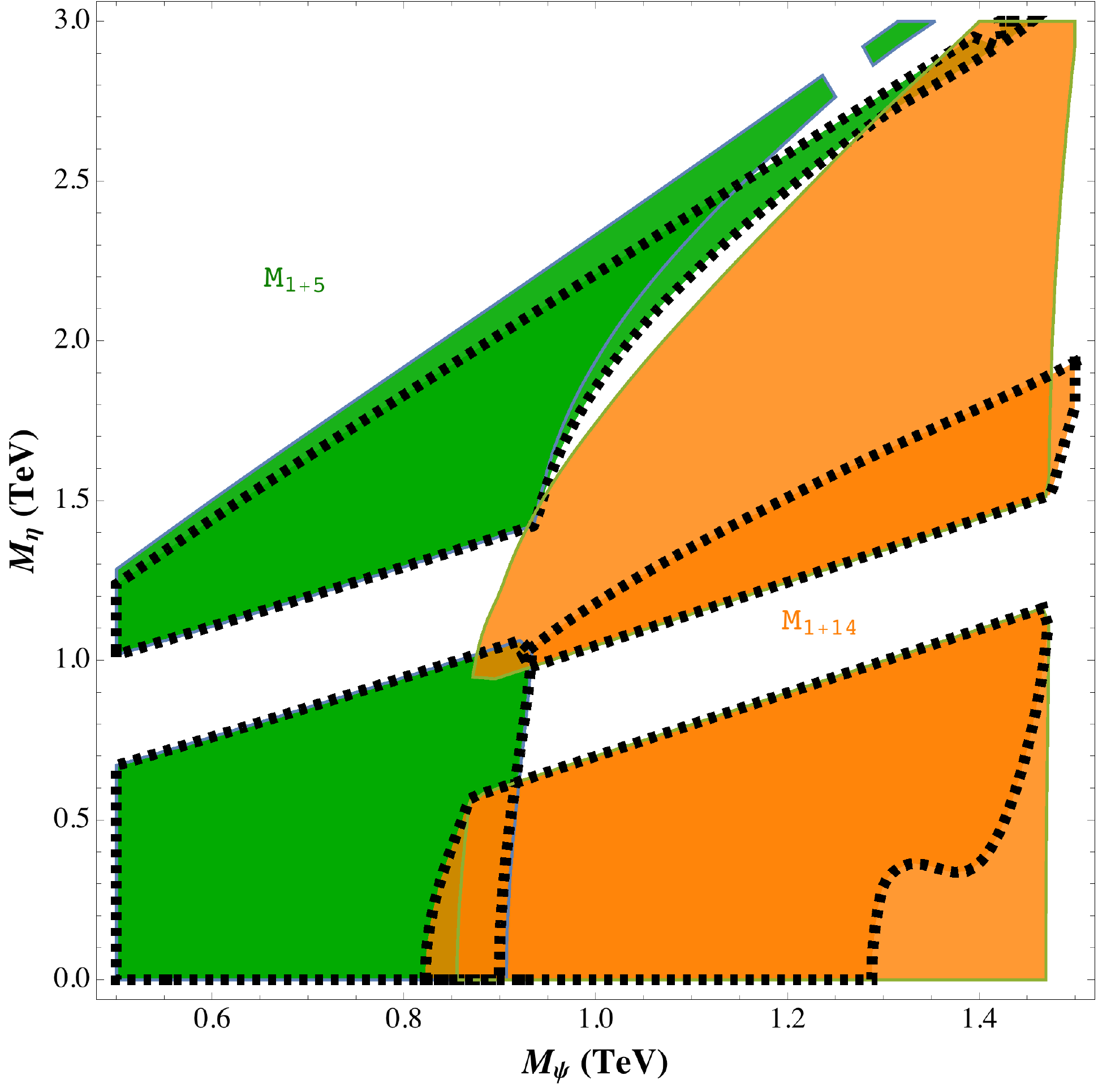}
\caption{\sf Parameter space $\left(M_\Psi,\,M_\eta\right)$ by fixing $\xi=0.2$ at the fourplet (left) and singlet models (right). Two situations have been scanned, $\alpha=0,1$ (thick-dashed border) by imposing recent bounds~\cite{Sirunyan:2017pks} on top partner searches through top-like decays into $b W$ final states at all models. }
\label{MPsi-Meta}
\end{center}
\end{figure}

The previous scalar mass ranges have been tested in previous analysis.  Searches for a Higgs-like bosons, decaying into long-lived exotic particles, have yielded no excess above the background expectation~\cite{Aaij:2016isa} in the range 80-140 TeV. Likewise, experimental evidences for a massive long-lived particles decaying semileptonically in the LHCb have not been found at the EW scale~\cite{Aaij:2016xmb}. Additionally, heavy scalar resonances decaying into $W W$ in the $e\nu \mu \nu$ final state, via $pp$ collisions at $\sqrt{s} = 13$ TeV with an integrated luminosity of $36.1\,\text{fb}^{-1}$, have not been detected either, and with no significant excess of events beyond the SM background prediction in the mass range 200-5000 TeV, following the searches performed by the ATLAS Collaboration~\cite{Aaboud:2017gsl}. Despite having no experimental evidence for Higgs-like particles at EW scale nor heavy resonance at the TeV range, the allowed range masses in Fig.~\ref{Meta-Xi} give us a clue for the mass regimes to be explored in future facilities and experiments in the shed light of exotic top partners decays and their interactions with possible new scalar matter in the nature. 

\nt Finally, a fuller understanding of the parametric dependence is gained by fixing the EW and GB scales while simultaneously scanning the top partner and the scalar mass scales. Fig.~\ref{MPsi-Meta} gathers the allowed areas for $\xi=0.2$. Whether the extra fermion-resonance couplings proposed here are considered or not, the region from $\fA$ entirely falls inside the exclusion limit of~\cite{Sirunyan:2017pks}, whilst the influence of $\LL_{\bf M\,+\,\eta}$ tends to diminish the permitted scalar range mass below $\sim$\, 2 TeV at $\fB$. A range for the scalar mass is correspondingly forbidden at both fourplet and singlet models. Disregarding extra derivative couplings, the exclusion bounds of~\cite{Sirunyan:2017pks} allow the higher scalar mass range $\sim$\, 1.8-3.0\,TeV at $\oA$, whereas $\sim$\, 0-1\,TeV and $\sim$\, 1.4-3\,TeV at $\oB$. Turning the extra interactions on will rule out part of the parameter space reducing it to $M_\eta\sim$\, 0.4-1\,TeV and $M_\eta\sim$\, 1.4-1.9\,TeV for $M_\Psi\sim$\, 1.42-1.5\,TeV. In this sense, the additional interactions encoded by $\LL_{\bf M\,+\,\eta}$ in~\eqref{Derivative-couplings} will be enlightening in refining further the mass range for the new scalar resonance in the theory, as well as the top partner mass scale emerging in this CHM scenarios.

\section{Summary}
\label{Summary}

\nt The interplay among three matter sectors is explored here: elementary, top partners and scalar resonances in a $SO(5)$ composite Higgs scenario. The scalar resonance $\eta$, here assumed to be a singlet of $SU(2)_L\times SU(2)_R$ is coupled to the $SO(5)$-invariant fermionic currents of Table~\ref{Fermion-currents-set}, here provided by the first time. The $SO(4)$ fourplet $\fourplet$ and singlet $\singlet$ source the top partners content of the model. Such matter framework spans four models in~\eqref{Models}, derivatively coupled to the $\eta$-resonance via the prescription~\eqref{Derivative-couplings} and subsequently explored through their involved parametric dependence.

Heavy scalar production and its decays have been extensively analysed along the benchmark range $M_\eta\sim [0.6,\,3]\,\text{TeV}$  and for a given model parameters election in Fig.~\ref{eta-Production-cross-sections}. The $\eta$-production is slightly altered when the fermion-resonance current interactions of~\eqref{Derivative-couplings} are included. The scalar heavy resonances is predominantly yielded at the model $\fA$, reaching rough cross section values of $\sim 150$ pb (0.1 pb) at $M_\eta \sim 0.6$ TeV (3 TeV) for $\xi=0.2$. Higher $\xi$-value enhances all the productions. Whether the elementary fermions are $\bf{5}$ or $\bf{14}$-embeddings, the fourplet scenario favours higher production values rather than the singlet one. We have tested two situations $\alpha=0,1$, with the aim of exploring the impact of the proposed additional couplings upon the implied phenomenology.

For the scalar decays, no extra fermion-resonance interactions entail dominant dijet, top-pair and gauge, Higgs channels, while subdominant single-double partners final states. The dijet channel is the dominant one for $M_\eta\lesssim 2$ TeV, becoming subdominant with respect the $W$-pair for a higher mass value. The scenario is altered after switching extra fermion-resonance couplings on. Certainly, those channels qualitatively diminish, being still the dominant ones, with a notorious enhancement for all the single-double partner final states in turn. The enhancement occurring at the partner final states, may be a slight departure, as in the case of the mode $b\cB$, or even an increase of one or two orders of magnitude roughly for the $t\Xtt$ and $\cT\cT$ channels.

Furthermore, the production of single-double partner final states has been scanned along the partner mass scale $M_\Psi$, being controlled by the model-dependent couplings $g_{ff \eta},\,g_{f_L f_L \eta},\,g_{f_R f_R \eta},\,g_{XX\eta}\,(g_{Xf\eta}),\,g_{X_L X_L \eta}\,(g_{X_L f_L \eta}),\,g_{X_R X_R\eta}\,(g_{X_R f_R\eta})$ along~\eqref{Scalar-Yukawa-derivative-couplings}-\eqref{Partners-Partners-eta} 
for the double\,(single) production, and by the analogous ones involving the gluon and the SM gauge fields. QCD drives the double production, as well as SM gauge, Higgs, and the $\eta$-mediated processes. Non-zero parametric-dependent modifications are induced as long as extra fermion-resonance effects are considered (Figs.~\ref{Double-partner-production}-\ref{Single-partner-production}). Slight enhancements occurs for the double production at $\fA$, whereas vanishing-tiny contributions are induced at the rest of the models due to the implied $f$-suppressed derivative couplings of~\eqref{Derivative-couplings}. $\cT\cT$ and $\cB\cB$ are dominantly produced in $pp$ collisions at $\fB$ as the involved quark partner masses are smaller than the corresponding ones at $\fA$. $\Xtt\Xtt$ does not distinguish the elementary embeddings representation as the involved partner masses are equal at both models. Generically, producing pairs either of $\Xtt$ or $\Xft$ will be kinematically favoured with respect to the double production of both $\cT$ and $\cB$, as their relatively higher masses are implied. Likewise, the pair production of the singlet $\widetilde{\cT}$ is dominant at $\oB$, as the involved masses result smaller at $\bf{14}$-elementary embeddings compared with the one at $\bf{5}$-scenario. On the other hand, cross section values are generically increased at the channels $t\Xtt$ and $t\widetilde{\cT}$ for the models $\fA$ and $\oB$ respectively if the extra couplings are included. Despite the absence of the flavour-changing neutral couplings in the charge −1/3 sector and of the $\cB \to hb$ channel at $\fA$, the final state $b\cB$ is still possible at the fourplet models via derivative couplings of~\eqref{Derivative-couplings}.

Finally, implementing the recent LHC searches for vector-like quarks production in $pp$-collisions at 13 TeV, we were able to exclude regions of the parameter space implied by our scenario (Figs.~\ref{MPsi-Xi}-\ref{MPsi-Meta}). If the scalar resonance $\eta$ is disregarded from the final states, part of the parameter spaces are permitted then at $\fB$ and at the singlet models, independently of the extra fermion-scalar interactions. By including the scalar field into the final states, a strongly bounded region, further constrained if the extra interactions in~\eqref{Derivative-couplings} are included, remains at $\fB$ and $\oA$. In this sense, those extra couplings are useful in discerning models and refining further their involved parameter space. The ranges $M_\eta\sim$\, 1.2-1.4\,TeV and $M_\eta\sim$\, 0.9-1.2\,TeV are forbidden at the $\bf{14}$-embedding scenarios, with $\oA$ favouring the higher scalar mass range 2.8-3\,TeV and with consistent $\xi$-values still possible. The allowed range masses in Fig.~\ref{Meta-Xi} give us a hint for the mass regimes to be explored in future experiments and in the shed light of exotic top partners and scalar decays in the nature. The higher scalar mass range 1.8-3\,TeV at $\oA$, whereas 0-1\,TeV and 1.4-3\,TeV at $\oB$ are allowed if no additional interactions are considered. Turning them on, will reduce the parameter space to $M_\eta\sim$\, 0.4-1\,TeV and $M_\eta\sim$\, 1.4-1.9\,TeV for $M_\Psi\sim$\, 1.4-1.5\,TeV. In this sense, the additional interactions encoded by $\LL_{\bf M\,+\,\eta}$ in~\eqref{Derivative-couplings} will be enlightening in refining further the mass range for the new scalar resonance in the theory, as well as the top partner mass scale emerging in this CHM scenarios.

Disregarding-selecting models and refining further their involved parameter dependence might be properly performed through the extra fermion-resonance couplings proposed here. Future observations of heavy scalars and top partners, as well as their subsequent posterior decays, might be clearly tackled via the effective Composite Higgs approach presented here. The interactions encoded by~\eqref{Derivative-couplings} and Table~\ref{Fermion-currents-set} would determine the scenario and the strength for the entailed effective terms by comparing their predictions with the experimental signals emerging at higher energies. The framework provided here represents a powerful tool in dealing with EFT approaches beyond SM scenarios, specifically in facing new interactions that might underlie the existence of exotic matter content in our nature, and hopefully discoverable at future high energy facilities.
\section*{Acknowledgements}

\nt J.Y. acknowledges valuable comments from  Felipe J. Llanes-Estrada, Juan J. Sanz-Cillero,  Dom\`{e}nec Espriu, Christophe Grojean and Oleksii Matsedonskyi, as well as the hospitality at the Universidad Complutense de Madrid, the Universitat de Barcelona and DESY where part of this work has been carried out. J.Y. thanks  the support of Fondecyt (Chile) grant No. 3170480. The work of A.Z. is supported by Conicyt (Chile) grants ACT1406 and PIA/Basal FB0821, and by  Fondecyt (Chile) grant 1160423. 
\appendix
\small

\section{CCWZ formalism}
\label{CCWZ}

\nt The $SO(4) \simeq SU(2)_L \times SU(2)_R$ unbroken generators and the broken ones parametrizing the coset $\textrm{SO}(5)/SO(4)$ in the fundamental representation are
\be
(T^a_\chi)_{IJ} = -\frac{i}{2}\left[\frac{1}{2}\varepsilon^{abc}
\left(\delta_I^b \delta_J^c - \delta_J^b \delta_I^c\right) \pm
\left(\delta_I^a \delta_J^4 - \delta_J^a \delta_I^4\right)\right],\qquad
T^{i}_{IJ} = -\frac{i}{\sqrt{2}}\left(\delta_I^{i} \delta_J^5 - \delta_J^{i} \delta_I^5\right)\,,
\label{eq:SO4_gen-SO5/SO4_gen}
\ee

\nt with $\chi= L,\,R$ and $a= 1,2,3$, while $i = 1, \ldots, 4$. The normalization of $T^{A}$'s is chosen as ${\rm Tr}[T^A, T^B] = \delta^{AB}$. The $4\times 4$ matrices $\tau^a$ appearing in the bilinear fourplets at $\fA$ and $\fB$ in Table~\ref{Fermion-currents-set} are defined as
\be
\left[T^{a},T^{i}\right]=\left(t^{a}\right)_{{j}{i}}T^{j}\,.
\label{taus}
\ee

\nt Gauging the SM subgroup of $SO(5)$ requires us to introduce local transformations via $U$ matrices that will couple the SM gauge fields to the composite resonances. The CCWZ $d$ and $e$ symbols are in order to do so 
\be
-U^t [A_\mu + i \partial_\mu ] U = d_\mu^{\hat a} T^{\hat a} + e^{a}_{\mu} T^a + e^{X}_{\mu}
\label{d-e}
\ee

\nt where $A_{\mu}$ stands for ${\cal G}_{\text{SM}}$ gauge fields 
\be
\begin{aligned}
A_\mu &= \frac{g}{\sqrt{2}}W^+_\mu T^-_L +\frac{g}{\sqrt{2}}W^-_\mu T^+_L+ g \left(\cw Z_\mu+\sw A_\mu \right)T_L^3+g' \left(\cw A_\mu-\sw Z_\mu \right)(T_R^3+Q_X)
\label{gfd}
\end{aligned}
\ee

\nt with $T^\pm_\chi=\left(T^1_\chi\mp i T^2_\chi\right)/\sqrt{2}$, the implied notation $(\cw,\,\sw)\equiv(\cos\theta_{\rm w},\,\sin\theta_{\rm w})$, and the SM couplings of $SU(2)_L$ and $\textrm{U}(1)_Y$, $g$, $g'$ respectively, where $Q_X$ is the $X$-charge matrix. The definition~\eqref{d-e} can be expanded in fields as
\be
d_\mu^i=\frac{\sqrt{2}}{f}(D_\mu h)^i+{\cal O} (h^3),\qquad
e^{a}_\mu = -A^{a}_\mu-\frac{i}{f^2}(h{\lra{D}_\mu}h)^a+{\cal O} (h^4),\qquad
e^{X}_\mu = -g^{\prime}  Q_X B_\mu \,,
\label{CCWZ-symbols}
\ee

\nt with $B_{\mu}$ the $U(1)_Y$ gauge boson. Covariant derivatives acting on the composite sector fields are built out of $e$ symbols. For the $\Psi$ field transforming in the fundamental representation of $SO(4)$ one has
\be
\nabla_\mu\Psi \,=\,D_\mu\Psi+i\,e_{\mu}^at^a\Psi\,.
\label{covariant-derivative}
\ee
The term $\slashed{e}=e_\mu\gamma^\mu$ is included in $\LL_{\text{comp}}$ to fully guarantee the $SO(5)$ invariance. Strength field tensors are straightforwardly introduced as
\be
e_{\mu \nu} = \partial_{\mu} e_{\nu} - \partial_{\nu} e_{\mu} + i g_{\rho} [e_{\mu},e_{\nu}], \qquad\qquad
e_{\mu \nu}^{X} = \partial_{\mu} e^X_{\nu} - \partial_{\nu} e^X_{\mu}.
\ee

\nt Finally, the covariant derivatives $D_\mu$ associated to each one of the elementary fields as well to the corresponding top partner are given by

\bea
D_\mu\,q_L&=&\left(\partial_\mu-ig W_\mu^i {\sigma^i\over 2}-i{1\over 6}g' B_\mu-i\,g_SG_\mu\right)q_L\,, \\
D_\mu\,u_R&=&\left(\partial_\mu-i{2\over 3}g' B_\mu-i\,g_SG_\mu\right)u_R \,,\\
D_\mu\fourplet &=& \left(\partial_\mu-i {2\over 3} g' B_\mu -i\,g_SG_\mu\right)\fourplet\,.
\label{cder}
\eea

\nt with $g,g'$ and $g_S$ the ${\textrm{SU}}(2)_L\times {\textrm{U}}(1)_Y$ and ${\textrm{SU}}(3)_c$ gauge couplings. Notice the gluon presence in the last covariant derivative as the top 
partners form a color triplet.

\section{Physical fermion masses}
\label{Physical-fermion-masses}

\subsection{${\bf 5}$-plets embeddings}
\label{Physical-fermion-masses-5}

\nt Considering both the fourplet and singlet models simultaneously, the mass matrices for the top-like and bottom-like sectors become
\be
\left(
\begin{array}{cccc}
 0 & \frac{1}{2} \mathit{f}\,\xi\, y_L & \mathit{f} (1-\frac{\xi}{2}) y_L & -\frac{\mathit{f} \sqrt{\xi }\,\tilde{y}_L}{\sqrt{2}} \\ [3mm]
 -\frac{\mathit{f} \sqrt{\xi }\,y_R}{\sqrt{2}} & -M_{\bf{4}} & 0 & 0 \\ [3mm]
 \frac{\mathit{f} \sqrt{\xi }\,y_R}{\sqrt{2}} & 0 & -M_{\bf{4}} & 0 \\ [3mm]
 \mathit{f} \sqrt{1-\xi }\,\tilde{y}_R & 0 & 0 & -M_{\bf{1}} \\
\end{array}
\right),\qquad\qquad \left(
\begin{array}{cc}
 0 & \mathit{f} y_L \\
 0 & -M_{\bf{4}} \\
\end{array}
\right)
\label{Top-bottom-matrices-5}
\ee

\nt being defined in the fermion field basis $\left(t,\,\,\Xtt\,\,\cT,\,\,\widetilde{\cT}\right)^T$ and $\left(b,\,\,\cB\right)^T$ respectively. After diagonalization the physical masses are
\beq
\begin{aligned}
&m_t=\sqrt{\frac{\xi \left(\tilde{\eta }_L\,\tilde{\eta }_R M_{\bf{1}} -\eta _L\eta _R M_{\bf{4}}\right)^2}{2\left(\eta _L^2+1\right) \left(\tilde{\eta }_R^2+1\right)}}\,,\quad\quad
&&\quad m_{\widetilde{\cT}}=M_{\bf{1}} \sqrt{\tilde{\eta }_R^2+1}\,,\quad\quad 
&&m_\cB=M_{\bf{4}} \sqrt{\frac{\eta _L^2+1}{\tilde{\eta }_R^2+1}}\,,  \\[4mm]
&m_\cT=M_{\bf{4}}\sqrt{\eta _L^2+1} \,,\quad\qquad
&&m_\Xtt=m_\Xft=M_{\bf{4}}\,,\quad\quad 
&&  \\
\end{aligned}
\label{Masses-expanded-5}
\eeq

\nt where the parameters $\eta_{L(R)}$ are defined through
\be
\eta _{L(R)} \equiv  \frac{y_{L(R)} \mathit{f}}{M_{\bf{4}}},\qquad\qquad
\tilde{\eta}_{L(R)}\equiv  \frac{\tilde{y}_{L(R)} \mathit{f}}{M_{\bf{1}}}.
\label{eta-parameters}
\ee

\subsection{${\bf 14}$-plets embeddings}
\label{Physical-fermion-masses-14}

\nt Considering both the fourplet and singlet models simultaneously, the mass matrix for the top-like sector is
\be
\small{
\left(
\begin{array}{cccc}
 -\mathit{f} \sqrt{2\,(1-\xi)\,\xi }\,y_R & \frac{1}{2} \mathit{f} \left(2 \xi +\sqrt{1-\xi }-1\right) y_L & \frac{1}{2} \mathit{f} \left(-2 \xi +\sqrt{1-\xi }+1\right) y_L & -\mathit{f} \sqrt{2\,(1-\xi)\,\xi }\,\tilde{y}_L \\
 0 & -M_{\bf{4}} & 0 & 0 \\
 0 & 0 & -M_{\bf{4}} & 0 \\
 0 & 0 & 0 & -M_{\bf{1}} \\
\end{array}
\right)
\label{Top-matrix-14}
}
\ee

\nt while the one for the bottom-like sector becomes
\be
\left(
\begin{array}{cc}
 0 & \mathit{f} \sqrt{1-\xi }\,y_L \\
 0 & -M_{\bf{4}} \\
\end{array}
\right)
\label{Bottom-matrix-14}
\ee

\nt The corresponding physical masses are
\beq
\begin{aligned}
&m_t=\sqrt{2} \sqrt{\frac{\xi \left(M_{\bf{1}} \tilde{\eta }_L \tilde{\eta }_R+M_{\bf{4}} \eta _R\right)^2}{\left(\eta _L^2+1\right) \left(\tilde{\eta }_R^2+1\right)}}\,,\quad\quad
&&\quad m_{\widetilde{\cT}}=\frac{M_{\bf{1}}}{\sqrt{\tilde{\eta }_R^2+1}}\,,\quad\quad 
&&m_\cB=M_{\bf{4}} \sqrt{\frac{\eta _L^2+1}{\tilde{\eta }_R^2+1}}\,,  \\[4mm]
&m_\cT=M_{\bf{4}}\sqrt{\eta _L^2+1} \,,\quad\qquad
&&m_\Xtt=m_\Xft=M_{\bf{4}}\,,\quad\quad 
&&  
\end{aligned}
\label{Masses-expanded-14}
\eeq

\begin{figure}
\begin{center}\includegraphics[scale=0.32]{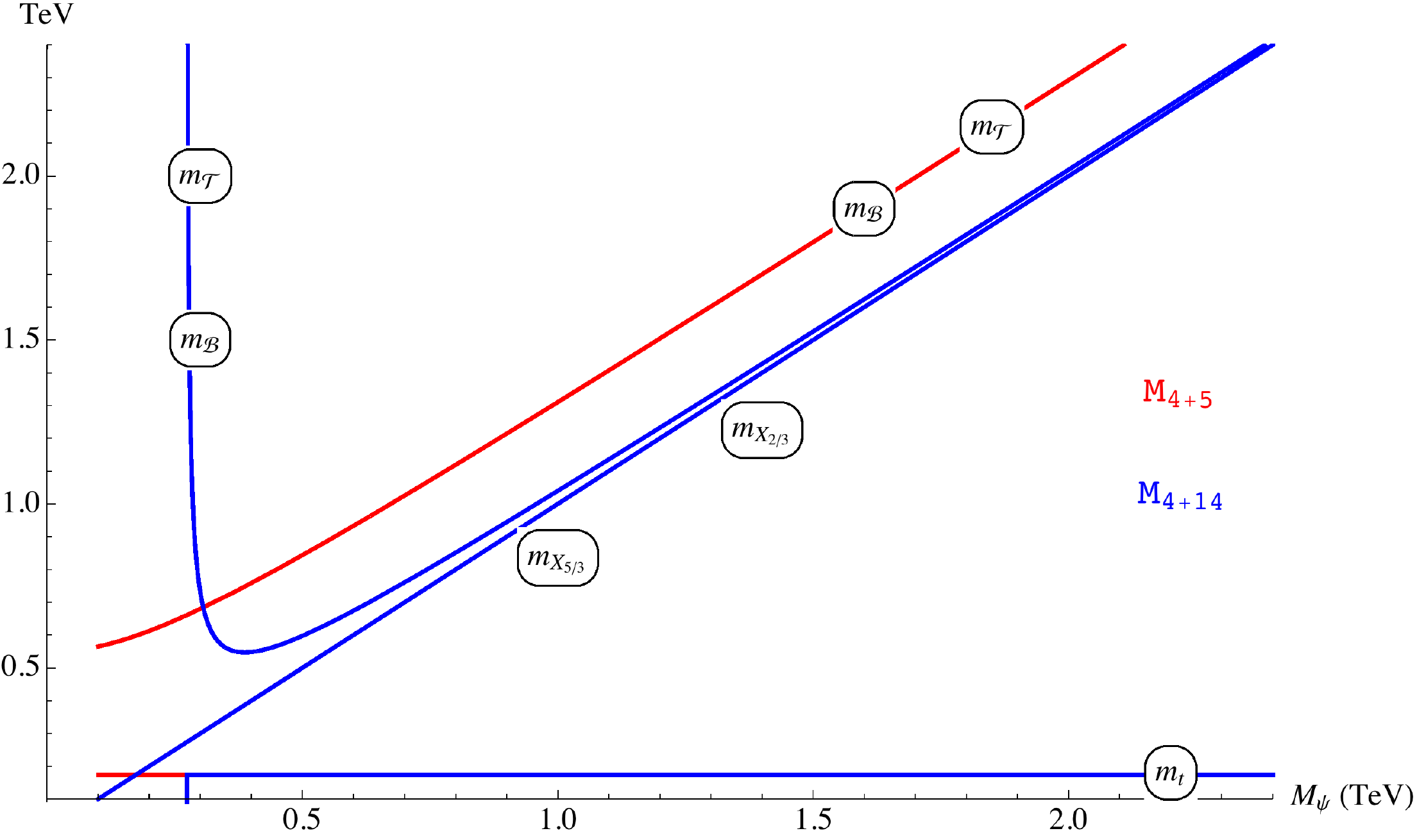}
\hspace*{1.5cm}
\includegraphics[scale=0.31]{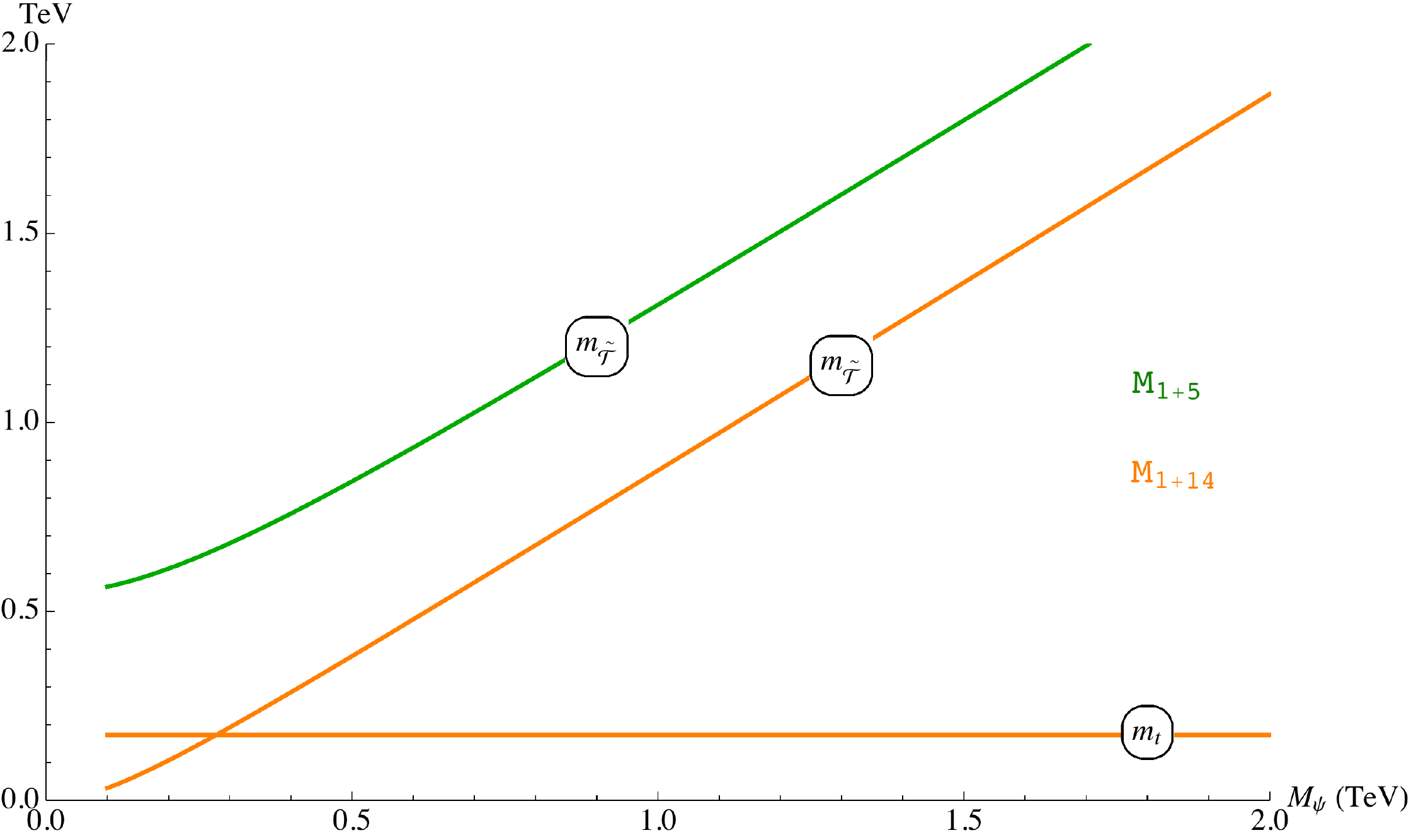}
\caption{\sf Spectrum of masses and their dependence on the NP scale $M_{\bf{4}}=M_{\bf{1}}=M_\Psi$ for the fourplet (left) and singlet cases (right), for $\xi=0.2$ and setting $\eta_L=\eta_R$, $\tilde{\eta }_L=\tilde{\eta }_R$.}
\label{4plet-singlet-masses}
\end{center}
\end{figure}

\section{Effective couplings}
\label{Effective-couplings}

\nt Here we report the model-dependent couplings $g_{ff \eta},\,g_{f_L f_L \eta},\,g_{f_R f_R \eta},\,g_{XX\eta},\,g_{X_L X_L \eta},\,g_{X_R X_R\eta}$ through~\eqref{Scalar-Yukawa-derivative-couplings} and~\eqref{Partners-Partners-eta} at the model $\fA$ as an example. To shorten the involved expressions, we restrict ourselves up to the zeroth order in $\xi$. Concerning the set $g_{ff \eta}$ and $g_{XX\eta}$ we have

\beq
\begin{aligned}
&g_{tt \eta}=0\,,\quad\quad
&&\quad g_{bb \eta}=\frac{\tilde{\eta }_R y_{q \psi }}{\sqrt{\left(\eta _L^2+1\right) \left(\tilde{\eta }_R^2+1\right)}}\,,\quad\quad 
&&g_{\cT\cT \eta}=-\frac{\eta _L y_{q \psi }}{\sqrt{\eta _L^2+1}}\,,  
\\[4mm]
&g_{\cB\cB \eta}=-\frac{\eta _L y_{q \psi }}{\sqrt{\left(\eta _L^2+1\right) \left(\tilde{\eta }_R^2+1\right)}} \,,\quad\qquad
&&\quad g_{\Xtt\Xtt \eta}=g_{\Xft\Xft \eta}=0.\quad\quad 
&&  \\
\end{aligned}
\label{Couplings-ffeta}
\eeq

\nt For the set $g_{f_L f_L \eta}$ and $g_{X_LX_L\eta}$ one has
\beq
\begin{aligned}
&g_{t_Lt_L \eta}=g_{b_Lb_L \eta}=\frac{\eta _L \left(\alpha _{\psi } \eta _L+2 \alpha _{q \psi }\right)+\alpha _q}{\sqrt{2} \mathit{f} \left(\eta _L^2+1\right)}\,,\quad\quad
&&\quad g_{\cT_L\cT_L \eta}=g_{\cB_L\cB_L \eta}=\frac{\alpha _{\psi }+\eta _L \left(\eta _L \alpha _q-2 \alpha _{q \psi }\right)}{\sqrt{2} \mathit{f} \left(\eta _L^2+1\right)}\,,\quad\quad 
\\[4mm]
&\quad g_{\Xtt_L\Xtt_L \eta}=g_{\Xft_L\Xft_L \eta}=\frac{\alpha _{\psi }}{\sqrt{2} \mathit{f}}.
&&\quad \qquad 
&&  \\
\end{aligned}
\label{Couplings-fLfLeta}
\eeq

\nt The degeneracy among some couplings are spoiled once higher $\xi$-order terms are considered. For $g_{f_R f_R \eta}$ and $g_{X_RX_R\eta}$ we have
\beq
\begin{aligned}
&g_{t_Rt_R \eta}=\frac{\alpha _u}{\sqrt{2} \mathit{f} \left(\tilde{\eta }_R^2+1\right)}\,,\quad\quad
&&\quad g_{b_Rb_R \eta}=\frac{\alpha _{\psi } \tilde{\eta }_R^2}{\sqrt{2} \mathit{f} \left(\tilde{\eta }_R^2+1\right)}\,,\quad\quad 
&&g_{\cT_R\cT_R \eta}=\frac{\alpha _{\psi }}{\sqrt{2} \mathit{f}}\,,  
\\[4mm]
&g_{\cB_R\cB_R \eta}=\frac{\alpha _{\psi }}{\sqrt{2} \mathit{f} \left(\tilde{\eta }_R^2+1\right)} \,,\quad\qquad
&&\quad g_{\Xtt_R\Xtt_R \eta}=g_{\Xft_R\Xft_R \eta}=\frac{\alpha _{\psi }}{\sqrt{2} \mathit{f}}.\quad\quad 
&&  \\
\end{aligned}
\label{Couplings-fRfReta}
\eeq

\nt Concerning the \emph{mixed} couplings $g_{Xf\eta},\,g_{X_L f_L \eta},\,g_{X_R f_R\eta}$ in~\eqref{Partners-fermion-eta}, they have similar analogous expressions and are not listed here for briefness reasons.


\providecommand{\href}[2]{#2}\begingroup\raggedright\endgroup

\end{document}

%% file: Fdiagrams/DoubleProduction-1.tex
\begin{fmffile}{DoubleProduction-1}
    \begin{fmfgraph*}(80,60)
        \fmfleft{i1,i2}
        \fmfright{o1,o2}
        \fmf{gluon}{i1,v1}
        \fmf{fermion}{v1,o1}        
        \fmf{gluon}{i2,v2}
        \fmf{fermion}{o2,v2}        
        
        \fmfv{lab=$g$}{i1}
        \fmfv{lab=$g$}{i2}
        \fmfv{lab=$X$}{o1}   
        \fmfv{lab=$\bar{X}$}{o2}           
        \fmf{fermion,label=$X$}{v2,v1}                   
 
   \end{fmfgraph*}
\end{fmffile}

%% file: Fdiagrams/DoubleProduction-3.tex
\begin{fmffile}{DoubleProduction-3}
    \begin{fmfgraph*}(80,60)
        \fmfleft{i1,i2}
        \fmfright{o1,o2}
        \fmf{fermion}{i1,v1,i2}
        \fmf{fermion}{o1,v2,o2}        
        
        \fmfv{lab=$\bar{u},,\bar{d}$}{i1}
        \fmfv{lab=$u,,d$}{i2}
        \fmfv{lab=$X$}{o1}   
        \fmfv{lab=$\bar{X}$}{o2}           
        \fmf{gluon,label=$g$,label.dist=0.5cm}{v2,v1}                   
 
   \end{fmfgraph*}
\end{fmffile}

%% file: Fdiagrams/DoubleProduction-2.tex
\begin{fmffile}{DoubleProduction-2}
    \begin{fmfgraph*}(80,60)
        \fmfleft{i1,i2}
        \fmfright{o1,o2}
        \fmf{fermion}{i1,v1,i2}
        \fmf{fermion}{o1,v2,o2}        
        
        \fmfv{lab=$\bar{u},,\bar{d}$}{i1}
        \fmfv{lab=$u,,d$}{i2}
        \fmfv{lab=$X$}{o1}   
        \fmfv{lab=$\bar{X}$}{o2}           
        \fmf{dashes,label=$h,,\eta$}{v2,v1}                   
 
   \end{fmfgraph*}
\end{fmffile}

%% file: Fdiagrams/DoubleProduction-4.tex
\begin{fmffile}{DoubleProduction-4}
    \begin{fmfgraph*}(80,60)
        \fmfleft{i1,i2}
        \fmfright{o1,o2}
        \fmf{fermion}{i1,v1,i2}
        \fmf{fermion}{o1,v2,o2}        
        
        \fmfv{lab=$\bar{u},,\bar{d}$}{i1}
        \fmfv{lab=$u,,d$}{i2}
        \fmfv{lab=$X$}{o1}   
        \fmfv{lab=$\bar{X}$}{o2}           
        \fmf{photon,label=$V$}{v2,v1}                   
 
   \end{fmfgraph*}
\end{fmffile}

%% file: Fdiagrams/SingleProduction-1.tex
\begin{fmffile}{SingleProduction-1}
    \begin{fmfgraph*}(90,70)
        \fmfleft{i1,i2}
        \fmfright{o1,o2,o4}
        \fmf{gluon}{i1,v1}
        \fmf{fermion}{v1,o1}        
        \fmf{fermion}{i2,v2}
        \fmf{fermion}{o3,o2}
        \fmf{fermion}{o4,o3}      
        
        \fmfv{lab=$g$}{i1}
        \fmfv{lab=$u,,d$}{i2}
        \fmfv{lab=$d,,u$}{o1}   
        \fmfv{lab=$\bar{X}$}{o2} 
        \fmfv{lab=$q$}{o4}           
        \fmf{fermion,label=$d,,u$}{v2,v1}
        \fmf{photon,label=$W^\pm$,label.dist=-0.5cm}{v2,o3}                     
 
   \end{fmfgraph*}
\end{fmffile}

%% file: Fdiagrams/SingleProduction-3.tex
\begin{fmffile}{SingleProduction-3}
    \begin{fmfgraph*}(90,70)
        \fmfleft{i1,i2}
        \fmfright{o1,o2,o4}
        \fmf{fermion}{i2,v1}
        \fmf{gluon}{v1,i1}        
        \fmf{fermion}{v2,o4}        
        \fmf{fermion}{o2,o3,o1} 
                
        \fmfv{lab=$g$}{i1}
        \fmfv{lab=$u,,d$}{i2}
        \fmfv{lab=$q$}{o1}   
        \fmfv{lab=$\bar{X}$}{o2}           
        \fmfv{lab=$u,,d$}{o4}           
        \fmf{fermion,label=$u,,d$,label.dist=-0.5cm}{v1,v2}                   
        \fmf{photon,label=$W^\pm$,label.dist=0.2cm}{v2,o3}                           
 
   \end{fmfgraph*}
\end{fmffile}

%% file: Fdiagrams/SingleProduction-2.tex
\begin{fmffile}{SingleProduction-2}
    \begin{fmfgraph*}(80,60)
        \fmfleft{i1,i2}
        \fmfright{o1,o2}
        \fmf{fermion}{i1,v1,i2}
        \fmf{fermion}{o1,v2,o2}        
        
        \fmfv{lab=$\bar{u},,\bar{d}$}{i1}
        \fmfv{lab=$u,,d$}{i2}
        \fmfv{lab=$q$}{o1}   
        \fmfv{lab=$\bar{X}$}{o2}           
        \fmf{dashes,label=$h,,\eta$}{v2,v1}                   
 
   \end{fmfgraph*}
\end{fmffile}

%% file: Fdiagrams/SingleProduction-4.tex
\begin{fmffile}{SingleProduction-4}
    \begin{fmfgraph*}(80,60)
        \fmfleft{i1,i2}
        \fmfright{o1,o2}
        \fmf{fermion}{i1,v1,i2}
        \fmf{fermion}{o1,v2,o2}        
        
        \fmfv{lab=$\bar{u},,\bar{d}$}{i1}
        \fmfv{lab=$u,,d$}{i2}
        \fmfv{lab=$q$}{o1}   
        \fmfv{lab=$\bar{X}$}{o2}           
        \fmf{photon,label=$V$}{v2,v1}                   
 
   \end{fmfgraph*}
\end{fmffile}